\documentclass[flushrt]{emulateapj}

\usepackage{subfigure} 
\usepackage{epsfig}

\usepackage{natbib}

\usepackage{longtable}

\newcommand{\npar}{\par \vspace{2.3ex plus 0.3ex minus 0.3ex}} 

\usepackage{lscape}

\begin{document}
\title{Constraints on the low-mass end of the mass-metallicity relation at $z=1-2$ from lensed galaxies.} 
\author{Eva Wuyts\altaffilmark{1,2}, Jane R. Rigby\altaffilmark{3}, Keren Sharon\altaffilmark{2,4}, Michael D. Gladders\altaffilmark{1,2}}

\altaffiltext{1}{Department of Astronomy and Astrophysics, University of Chicago, 5640 S. Ellis Av., Chicago, IL 60637}
\altaffiltext{2}{Kavli Institute for Cosmological Physics, University of Chicago, 5640 South Ellis Avenue, Chicago, IL 60637}
\altaffiltext{3}{Observational Cosmology Lab, NASA Goddard Space Flight Center, Greenbelt, MD 20771}  
\altaffiltext{4}{Department of Astronomy and Astrophysics, University of Michigan}  

\begin{abstract}
We present multi-wavelength imaging and near-IR spectroscopy for ten gravitationally lensed galaxies at $0.9<z<2.5$ selected from a new, large sample of strong lens systems in the Sloan Digital Sky Survey (SDSS) DR7. We derive stellar masses from the rest-frame UV to near-IR spectral energy distributions, star formation rates (SFR) from the dust-corrected H$\alpha$ flux, and metallicities from the [N~II]/H$\alpha$ flux ratio. We combine the lensed galaxies with a sample of sixty star-forming galaxies from the literature in the same redshift range for which measurements of [N~II]/H$\alpha$ have been published. Due to the lensing magnification, the lensed galaxies probe intrinsic stellar masses that are on average a factor of 11 lower than have been studied so far at these redshifts. They have specific star formation rates that are an order of magnitude higher than seen for main-sequence star-forming galaxies at $z\sim2$. We measure an evolution of $0.16\pm0.06$~dex in the mass-metallicity relation between $z\sim1.4$ and $z\sim2.2$. In contrast to previous claims, the redshift evolution is smaller at low stellar masses. We do not see a correlation between metallicity and SFR at fixed stellar mass. The combined sample is in general agreement with the local fundamental relation between metallicity, stellar mass and SFR from Mannucci et al. (2010, 2011). Using the Kennicutt-Schmidt law to infer gas fractions, we investigate the importance of gas inflows and outflows on the shape of the mass-metallicity relation using simple analytical models. This suggests that the \cite{Maiolino2008} calibration of the [N~II]/H$\alpha$ flux ratio is biased high. We conclude that both an absolute metallicity calibration and direct measurements of the gas mass are needed to use the observed mass-metallicity relation to gain insight into the impact of gas flows on the chemical evolution of galaxies.
\subjectheadings{galaxies: high-redshift, galaxies: evolution, strong gravitational lensing}                                    
\end{abstract}

\section{Introduction}
Our current understanding of the build-up of a galaxy's stellar mass and metallicity over cosmic time involves a subtle interplay of gas accretion, star formation and galactic winds. The accretion of metal-poor gas from the intergalactic medium (IGM) fuels the star formation; subsequent generations of stars contribute metals to the surrounding interstellar medium (ISM) and power galactic-scale outflows of enriched gas. This simple picture of chemical evolution is generally accepted, but the detailed physical mechanisms behind each of its components are far from well understood. There is as of yet no consensus on how to incorporate physically accurate prescriptions of gas inflows and outflows into semi-analytical and numerical models of galaxy formation.

Observations have not been able to provide much guidance. In recent years, spectral line measurements have revealed that galactic-scale outflows are a ubiquitous feature of star-forming galaxies at all redshifts \citep{Shapley2003,Weiner2009,Steidel2010}. However, the extent to which these outflows affect the chemical evolution of galaxies depends on the mass outflow rate, which has only been measured for one high redshift source: \cite{Pettini2002} found a mass loss rate at least comparable to and possibly many times greater than the star formation rate for the highly magnified lensed galaxy MS1512-cB58 \citep{Yee1996,Williams1996}. It is unclear whether these outflows will escape the gravitational potential of their host galaxy or rain back down. The high metallicity of the diffuse IGM in galaxy groups and clusters is indirect evidence that at least some fraction of the metals in these winds is lost to the environment (e.g. Renzini et al. 1993). 

The accretion of metal-poor gas is generally thought to happen along filaments and is refered to as cold-stream accretion or cold flows. It has received much recent attention in numerical simulations of galaxy formation as the dominant mechanism which fuels the star formation of high-redshift galaxies (e.g. Dekel et al. 2009). The existence of massive stellar populations at $z\sim2$ with old ages, high star formation rates (SFR), and continuous or slowly decreasing star formation histories requires a replenishment of their gas reservoirs by accretion to sustain this vigorous, extended star formation \citep{Shapley2005,Papovich2006,Erb2008}. However, direct observations of cold flows are still lacking. Line profile measurements at $z\sim2-3$ have found little evidence of infalling material, though the expected narrow velocity range of inflowing gas can easily be disguised by the several hundred km/s spread observed in most outflows \citep{Steidel2010}. Additionally, the covering factor of filaments for each galaxy is small, such that projection effects strongly reduce the probability of detecting them in absorption \citep{Steidel2010}. The recent discovery of two extremely metal-poor Lyman limit systems at $z\sim3$ in the spectra of bright quasars presents the first observational hints of cold flows at high redshift \citep{Fumagalli2011}.
\npar
Until inflows and outflows can be observed directly and their mass flow rates can be quantified, measurements of gas-phase metallicities and their relation to stellar mass can provide indirect insights into the impact of these gas flows on the chemical evolution of galaxies over time. In the closed-box model of chemical enrichment, the metallicity uniquely depends on the stellar mass as a record of the accumulated conversion of gas into stars. Deviations from this closed-box model prediction point towards metal exchanges between the galaxy and its environment through gas flows. 
\cite{Lequeux1979} first noted the increase in metallicity for more massive galaxies. The correlation was quantified for the Sloan Digital Sky Survey (SDSS) by \cite{Tremonti2004}, who found a tight ($\pm0.1$~dex) mass-metallicity relation (MZR) with a relatively steep slope, which flattens above $10^{10.5}$~M$_\odot$. A study of dwarf galaxies has extended the relation to much lower stellar masses ($10^{6.5}$~M$_\odot$) with a similar low scatter \citep{Lee2006}. A clear evolution of the MZR with redshift has been found using galaxy samples at $0.4<z<1.0$ \citep{Savaglio2005}, $1.0<z<1.5$ \citep{Shapley2005,Liu2008}, $2.0<z<2.5$ \citep{Erb2006a}, and $z\sim3$ \citep{Maiolino2008,Mannucci2009}: at a fixed stellar mass the metallicity decreases at earlier times. 

Much effort is being invested in using both semi-analytic models and cosmological hydrodynamic simulations to explain the observed MZR. 
One common idea states that galactic winds 
can more  efficiently remove metals from the shallower potential wells of low-mass galaxies, thus reproducing the observed MZR (e.g. Kobayashi et al. 2007; Scannapieco et al. 2008, Spitoni et al. 2010). In an alternative scenario, low-mass galaxies convert their gas reservoirs into stars over longer timescales than more massive galaxies. They subsequently show larger gas-to-stellar mass fractions and are simply less enriched. Some studies have required galactic winds to regulate this mass-dependent star formation efficiency (e.g. Brooks et al. 2007, Finlator \& Dave 2008). In contrast, \cite{Tassis2008} employ a critical density treshold for the onset of star formation and consequently do not require outflows to reduce the star formation efficiency of low-mass galaxies. Finally, a net dilution of the gas-phase metallicity through inflows of metal-poor gas when the timescale for star formation falls below the accretion timescale can explain the observed MZR (e.g. Dalcanton et al. 2004).


Compared to this extensive theoretical effort, observational studies of the MZR are much less developed. Beyond $z\gtrsim 1$, metallicity measurements of individual galaxies are still fairly rare. A large investment of near-IR spectroscopy is required to detect the rest-frame optical nebular emission lines used as metallicity indicators. Faint, low-mass galaxies are even more expensive to observe, limiting current $z\sim1-2$ studies to stellar masses $\ge10^{9.5}$~M$_\odot$. Gravitational lensing provides unique opportunities to extend metallicity studies to lower stellar masses. Strongly lensed galaxies are typically magnified by factors of a few up to $>50$ and probe systems that are intrinsically an order of magnitude less massive with significantly increased observing efficiency.

Only a handful of the known lensed galaxies in the literature have robust measurements of stellar mass and metallicity \citep{Teplitz2000, Lemoine2003, Hainline2009, Bian2010}. Many lensed galaxies are discovered serendipitously, and often not much follow-up is invested beyond the discovery paper. Recently \cite{Richard2011} presented metallicity measurements for $\sim$15 gravitationally lensed sources from the Hubble Space Telescope (HST) archive. To create a large sample of highly magnified lensed galaxies, we have systematically surveyed the galaxy clusters in the Sloan Digital Sky Survey (SDSS) Data Release 7 and identified a uniformly selected, complete sample of $\sim$150 confirmed lens systems (Bayliss et al. 2011, M.~D. Gladders et al.~2012, in preparation). With an average $g$-band magnitude of m$_{AB}$=22.0, the lensed sources are sufficiently bright for multi-wavelength photometric and spectroscopic follow-up. 
The stellar populations and physical conditions of two galaxies from this sample at $z=2.76$ and $z=2.92$ have been studied extensively with multi-wavelength photometry and near-IR spectroscopy \citep{me2012}. In this paper we present similar observations for an additional ten lensed galaxies at $0.9<z<2.5$. We compare our results to measurements from the literature to extend the MZR for individual galaxies at $z=1-2$ down to lower stellar masses.

The paper is organized as follows. \S\ref{sec:data} describes our sample selection, observations and data reduction. We derive the galaxy properties - stellar mass, star formation rate, and metallicity - in \S\ref{sec:param}. \S\ref{sec:addata} summarizes the additional data taken from the literature and \S\ref{sec:results} presents the combined results. We discuss observational constraints on the origin of the MZR in \S\ref{sec:disc}. Throughout the paper we adopt a flat cosmology with $\Omega_M = 0.3$ and H$_0 = 70$\,km\,s$^{-1}$\,Mpc$^{-1}$. We assume a Chabrier initial mass function \citep{Chabrier2003} and quote all magnitudes in the AB system. 

\begin{deluxetable*}{ccccccccc}
\scriptsize
\tablewidth{0pc}
\tablecaption{Photometry \label{tab:phot}}
\tablehead{\colhead{source} &  \colhead{$g$} & \colhead{$r$}  & \colhead{$i$} & \colhead{$z$} & \colhead{$J$} & \colhead{$H$} & \colhead{3.6~\micron} & \colhead{4.5~\micron}} 
\startdata
SGAS J085137$+$333114 & 21.66$\pm$0.06 & 21.40$\pm$0.05 & 21.04$\pm$0.05 &         &         &         & 19.13$\pm$0.12 & 18.87$\pm$0.12 \\
SGAS J091538$+$382658 & 22.21$\pm$0.08 & 21.95$\pm$0.07 & 21.52$\pm$0.07 & 21.44$\pm$0.09 & 20.71$\pm$0.30 &         & 18.68$\pm$0.17 &        \\
SGAS J103842$+$484926 & 21.26$\pm$0.07 & 20.66$\pm$0.06 & 20.37$\pm$0.06 &         &         & 17.64$\pm$0.12 & 16.83$\pm$0.15 & 16.16$\pm$0.15 \\
SGAS J103844$+$484914 & 21.27$\pm$0.11 & 20.74$\pm$0.11 & 20.02$\pm$0.11 &         &         & 18.67$\pm$0.31 & 17.52$\pm$0.21 & 17.08$\pm$0.21 \\ 
SGAS J111124$+$140901 & 21.75$\pm$0.11 & 21.57$\pm$0.11 & 21.87$\pm$0.11 &         &         &         & 21.12$\pm$0.21 & 21.06$\pm$0.21 \\
SGAS J113809$+$275445 & 21.75$\pm$0.07 & 21.24$\pm$0.07 & 21.05$\pm$0.07 &        &         &        & 20.19$\pm$0.16 & 19.95$\pm$0.15 \\ 
SGAS J113809$+$275439 & 21.31$\pm$0.11 & 21.01$\pm$0.11 & 20.62$\pm$0.11 &         &         &         & 20.52$\pm$0.24 & 20.71$\pm$0.23 \\
SGAS J120924$+$264052 & 20.96$\pm$0.06 & 20.67$\pm$0.06 & 20.02$\pm$0.06 &        &         & 19.64$\pm$0.17 & 19.32$\pm$0.15 & 19.69$\pm$0.15 \\
SGAS J134334$+$415509 & 20.46$\pm$0.08 & 20.27$\pm$0.06 & 20.39$\pm$0.07 & 20.30$\pm$0.07 & 19.96$\pm$0.14 &        & 19.32$\pm$0.15 &         \\ 
SGAS J142039$+$395454 & 21.11$\pm$0.07 & 20.63$\pm$0.06 & 20.27$\pm$0.06 &         &         &         & 18.75$\pm$0.23 & 18.61$\pm$0.23 \\ 
\enddata
\tablecomments{Magnitudes are quoted in the AB system.} 
\end{deluxetable*}

\section{Sample Selection, Observations and Data Reduction}
\label{sec:data}
The lensed galaxies discussed in this paper are drawn from the subset of 26 SDSS lens systems that have been targeted by the Gemini Multi-Object Spectrograph (GMOS) to measure velocity dispersions and dynamical masses for the lensing clusters \citep{Bayliss2011}. This effort has identified 69 unique lensed background sources with redshifts as high as $z=5.2$. We obtained near-IR spectra for a subset of 10 lensed galaxies within 8 of the lens systems, mostly targeting the primary giant arc in each system. These sources were selected from the larger sample based on observability, rest-frame UV surface brightness and the size of the arc, so as to take optimal advantage of the 42\arcsec\ long Keck/NIRSPEC slit. Throughout the paper we adopt the naming convention introduced in \cite{Koester2010}, which designates lensed sources as SGAS JHHMMSS+DDMMSS. 

\subsection{Imaging}
\label{subsec:im}
The optical imaging of our sample consists of $gri$ GMOS pre-imaging obtained to facilitate mask design for the GMOS spectroscopy presented in \cite{Bayliss2011}. Each image is a stack of two dithered exposures of 150~s each, taken with the detector binned $2\times2$ for a scale of 0.1454\arcsec~pixel$^{-1}$ and reduced using the Gemini IRAF\footnotemark[1] package. SGAS J091538$+$382658 and SGAS J134334$+$415509 have additional 300~s GMOS $z$-band data, obtained for the study of two lensed Lyman-$\alpha$ emitters at $z\sim5$ discovered in these fields \citep{Bayliss2010}. 
\footnotetext[1]{IRAF (Image Reduction and Analysis Facility) is distributed by the National Optical Astronomy Observatories, which are operated by AURA, Inc., under cooperative agreement with the National Science Foundation}

We have carried out near-IR imaging for three lensed galaxies with the Near-Infrared Camera and Fabry-Perot Spectrometer (NIC-FPS) on the 3.5~m telescope at the Apache Point Observatory (APO). SGAS J120924$+$264052 was observed for 3150~s in the $H$-band on 2008 May 19 UT, SGAS J091538$+$382658 was observed for 7200~s in the $J$-band on 2009 February 6 UT and SGAS J134334$+$415509 was observed for 5400~s in the $J$-band on 2009, March 16 UT. All data was taken under sub-arcsecond seeing conditions. A custom pipeline of standard IRAF tasks was used to sky-subtract and stack the dithered NIC-FPS images. 

Warm Spitzer observations with IRAC at 3.6 and 4.5\,$\mu$m were obtained for a larger sample of $\sim100$ lens systems through Spitzer program 70154 (PI: M.~D. Gladders). The lensed galaxies studied in this paper all have an integration time of 1200~s in each band, with the exception of SGAS J091538$+$382658 and SGAS J134334$+$415509. For these two sources, warm IRAC observations at 3.6\,$\mu$m were obtained through Spitzer program 60158 (PI: M.~D. Gladders) with a total integration time of 1800~s and 2700~s respectively \citep{Bayliss2010}. All IRAC data was reduced with the MOPEX software distributed by the Spitzer Science Center and drizzled to a finer pixel scale of $0\farcs6$\,pix$^{-1}$.
\npar
Photometry of the lensed galaxies is performed using largely the same method as described in detail in \cite{me2010}. In summary, we define ridgelines along the arcs and create non-circular photometric apertures as isophotes of the convolution of these ridgelines with the empirical PSF of the images. Final magnitudes are measured at an equivalent radius of twice the FWHM of the image and aperture-corrected to an equivalent radius of 6\arcsec\ based on the curve of growth of PSF reference stars. Aperture corrections for the IRAC data are applied as detailed in the IRAC handbook\footnotemark[2]; a different correction is made for point sources ($<8$\arcsec) and extended sources ($>8$\arcsec). We use GALFIT version 3.0 \citep{Peng2010} to mask neighboring galaxies that fall within the photometric apertures and nearby bright cluster galaxies with haloes that extend out to the location of the arc. This is especially important for the IRAC data due to the extended, non-circular PSF of the instrument. Final magnitudes are corrected for galactic extinction \citep{Schlegel1998} and presented in Table~\ref{tab:phot}. Photometric uncertainties include Poisson noise, absolute zeropoint uncertainties, uncertainties from the Spitzer aperture corrections as detailed in the instrument handbooks, and uncertainties from the masking of neighboring galaxies. This last effect often dominates the total uncertainty. 
\footnotetext[2]{http://ssc.spitzer.caltech.edu/irac/iracinstrumenthandbook/}

\subsection{Near-IR spectroscopy}
\label{subsec:spec}
\begin{figure*}
\center
\includegraphics[width=0.9\textwidth]{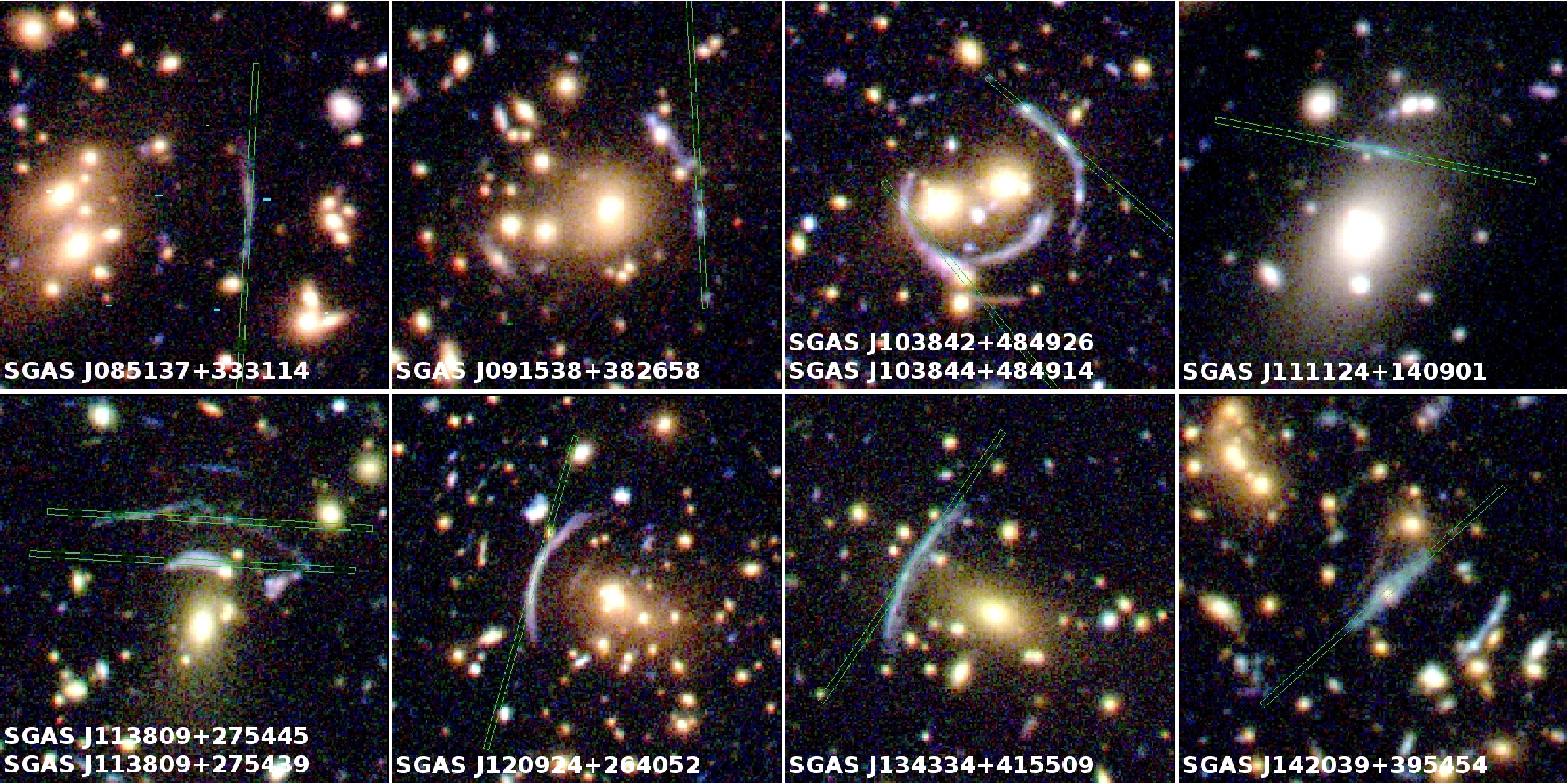}
\caption{Color-composite images of our lensed galaxies from the $gri$ GMOS pre-imaging, 50\arcsec\ on a side. North is up and East is left. The 0.76\arcsec\ $\times$ 42\arcsec\ NIRSPEC slit is overplotted for each source. For simplicity we show the slit centered on the source; in fact the source was placed on the left half and then the right half of the slit for the ABBA nod pattern, with the nods separated by 10-15\arcsec. \label{fig:finders}}
\end{figure*}

We obtained near-IR spectra for this sample using the NIRSPEC spectrograph \citep{mclean} on the Keck II telescope on the following nights of 2011: March 19 UT (full night), March 22 UT (half night), and March 23 UT (half night). Cirrus was present on the second and third nights. The seeing was measured at telescope re-focusing, and monitored from the CFHT seeing monitor; the measured seeing was highly variable, ranging from 0.7\arcsec\ to 2.4\arcsec. We used NIRSPEC's low-resolution mode with the 0.76\arcsec\ $\times$ 42\arcsec\ slit, and set the position angle for each source to maximize coverage of the extended arc. Figure~\ref{fig:finders} shows the location of the NIRSPEC slit on composite $gri$ images of the lensed galaxies (created from GMOS pre-imaging discussed in \S\ref{subsec:im}). Each source was acquired by offsetting from the brightest cluster galaxy on the near-IR slit-viewing camera; source acquisition was verified by direct imaging with this camera. During spectroscopic integration, the observations were nodded along the slit in an ABBA pattern, with typical exposure times of 600~s. To enable telluric calibration and fluxing, immediately before or after each target an A0V star was observed, ideally located less than 10 degrees from the target and with a similar airmass ($\delta (secz) < 0.1$). Table~\ref{tab:obslog} summarizes the spectroscopic observations.

\begin{deluxetable*}{lrllllll}
\scriptsize
\tablewidth{0pc}
\tablecaption{Spectroscopic Observation Log\label{tab:obslog}}
\tablehead{
\colhead{source} &  \colhead{PA}  & \colhead{filter} & \colhead{grating} & \colhead{$\lambda$}  
& \colhead{t} & \colhead{A0V star} & \colhead{extraction} \\
\colhead{} &  \colhead{}  & \colhead{} & \colhead{angle} & \colhead{[\micron]}  
& \colhead{[s]} & \colhead{} & \colhead{method}}
\startdata
SGAS J085137$+$333114 & 357 & N-6 & 33.50 & 1.60-2.02 & 1800 & HD71906  & each exposure  \\
SGAS J091538$+$382658 &  5   & N-5 & 36.72 & 1.51-1.79 & 5400 & HD71906  & each exposure  \\
SGAS J103842$+$484926 & 49  & N-6 & 35.00 & 1.93-2.32 & 2400 & HD99966  & each exposure  \\
SGAS J103844$+$484914 & 40  & N-3 & 34.08 & 1.16-1.37 & 3600 & HD99966  & stacked nods \\
SGAS J111124$+$140901 & 77  & N-6 & 34.50 & 1.81-2.23 & 3000 & 69Leo    & each exposure  \\
SGAS J113809$+$275445 & 87  & N-5 & 36.40 & 1.47-1.74 & 6000 & HD105388 & stacked nods \\	  
SGAS J113809$+$275439 & 87  & N-3 & 34.08 & 1.17-1.38 & 2400 & HD105388 & each exposure  \\
SGAS J120924$+$264052 & 163 & N-4 & 35.23 & 1.30-1.59 & 3400 & HD105388 & each exposure  \\
SGAS J134334$+$415509 & 146 & N-6 & 34.50 & 1.81-2.23 & 2400 & HD109615 & each exposure  \\
SGAS J142039$+$395454 & 312 & N-6 & 34.50 & 1.80-2.22 & 2400 & HD116405 & each exposure  \\
\enddata
\tablecomments{Columns are source name, position angle of the slit, NIRSPEC filter name, cross-disperser angle, wavelength coverage, total integration time, A0V star used for telluric correction and extraction method.}
\end{deluxetable*}

\npar
The spectra were reduced with the \textit{nirspec\_reduce} package written by G.~D.~Becker. The reduction is described in more detail in \citet{Rigby2011} and can be summarized as using lamp exposures to flatten the data, using skylines to calibrate wavelengths, and performing sky subtraction following \citet{Kelson2003}. We employ two different extraction strategies, depending on target brightness. For an object with fairly bright emission lines, we measure the spatial profile from H$\alpha$ 
on each exposure and use this to optimally extract the spectrum from that exposure. For the faintest objects, we stack the 2D sky-subtracted exposures for the A and B nod positions, fit the spatial profile of H$\alpha$ and optimally extract the stacked spectrum for each nod position. For each object, the extraction method is listed in Table~\ref{tab:obslog}. When the emission is concentrated in individual ``knots'' within the slit, we separately extract and analyze the spectrum of each knot.  

\begin{figure*}
\includegraphics[width=\textwidth]{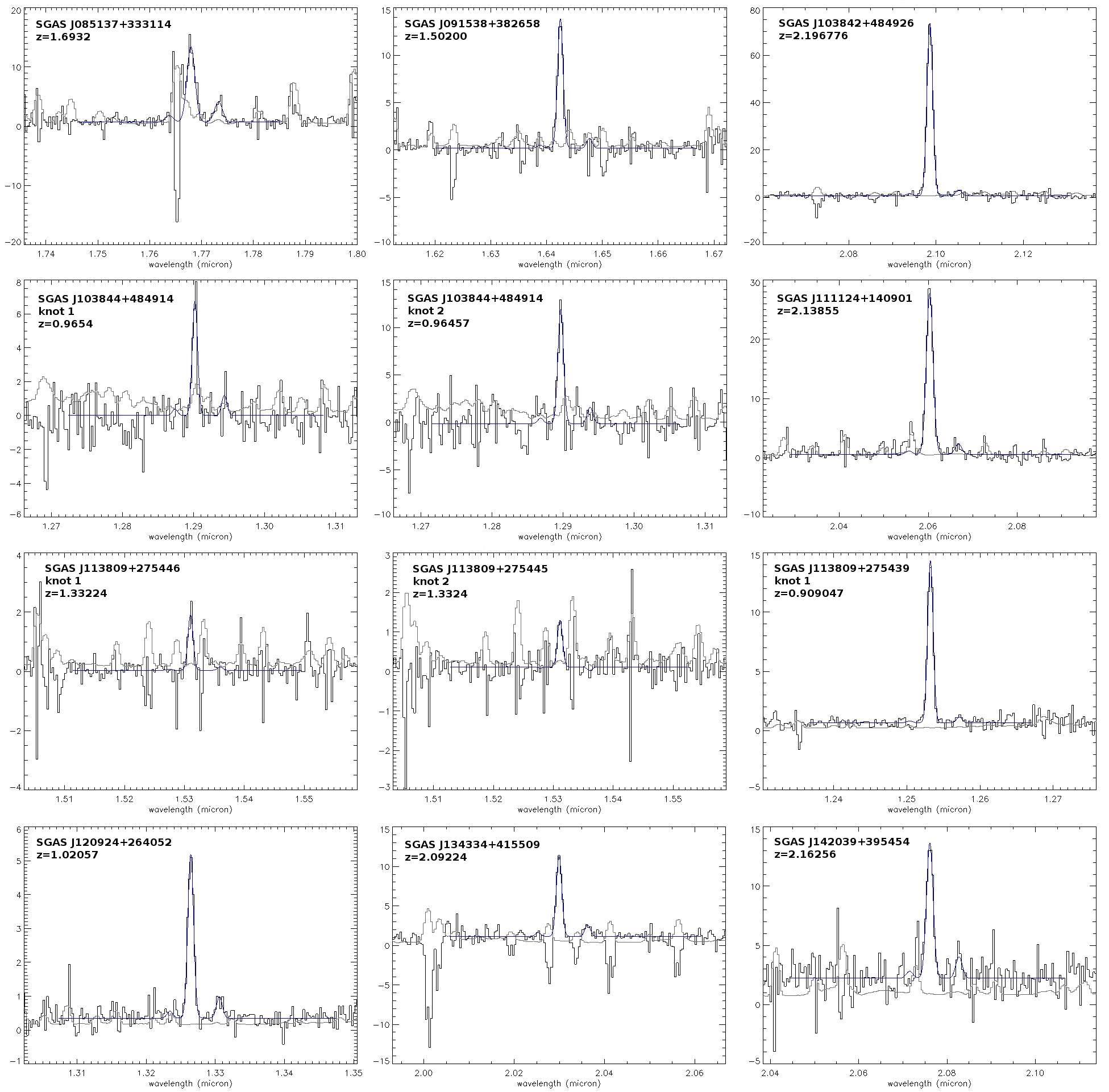}
\caption{Fluxed 1D NIRSPEC spectra of the lensed galaxies. The Y-axis shows flux in units of $10^{-17}$~erg~s$^{-1}$~cm$^{-2}$. Spectra are shown in black, 1~$\sigma$ error spectra in gray, and the best-fits to the H$\alpha$ and [N~II] emission lines in blue. The wavelength extent of each best-fit shows the region over which the continuum level was fit. \label{fig:spectra}}
\end{figure*}

Each extracted spectrum is corrected for telluric absorption and fluxed based on a A0V standard star spectrum, using the tool \textit{xtellcor\_general} \citep{Vacca2003}. Fluxing is thus appropriate for the fraction of the galaxy falling inside the slit, not for the whole galaxy. Given the highly variable seeing, the absolute fluxing may have significant systematic errors. For objects with a continuous arc morphology, we combine the individual flux-calibrated spectra from each exposure with a weighted average to produce a single fluxed spectrum, along with its 1~$\sigma$ error spectrum. When objects are composed of discrete emission ``knots'' or multiple images, we create a single fluxed spectrum for each knot or image in the same manner.
\npar
We derive fluxes for the H$\alpha$, [N~II] and [S~II] emission lines as follows using the IDL Levenberg-Marquardt least-squares code MPFITFUN \citep{mpfitfun}. First, to determine the redshift, we fit a single Gaussian to the H$\alpha$ emission; the uncertainty in the measured wavelength centroid is used in subsequent fitting. Second, we simultaneously fit three Gaussians to the system of the [N~II] emission lines and H$\alpha$, with the following physically motivated constraints:
\begin{itemize}
\item The wavelength centroids are set to the NIST rest wavelengths\footnotemark[3], redshifted by the redshift measured for H$\alpha$.  
\item Each wavelength centroid may vary by up to twice the 1~$\sigma$ wavelength uncertainty from the initial fit to H$\alpha$.
\item All lines must share a common linewidth that may vary freely. The initial linewidths are the best-fit values obtained from lines in the arc lamps.
\item The flux ratio of the [N~II] doublet is locked to the theoretical value computed by \citet{storey_zeippen00}.
\end{itemize}
\footnotetext[3]{http://www.pa.uky.edu/$\sim$peter/atomic/}
Finally, we measure the [S~II] emission by fixing the redshift and linewidth to the values obtained for the joint H$\alpha$ and [N~II] fit, and fitting the [S~II] doublet, allowing the flux ratio to vary freely. Table~\ref{tab:fluxes} reports the best-fit redshifts and flux measurements. Only [S~II] detections are included. For objects where individual ``knots'' or images were extracted separately, we report the results for each component, as well as the summed fluxes over all components, and the weighted mean redshift and its uncertainty. Figure~\ref{fig:spectra} shows the extracted spectra, the 1~$\sigma$ error spectra, and the best fit Gaussians to the [N~II] and H$\alpha$ emission lines. 

\begin{deluxetable*}{llllll}
\tabletypesize{\footnotesize}
\tablewidth{0pc}
\tablecaption{Measured redshifts and emission line fluxes. \label{tab:fluxes}}
\tablehead{
\colhead {source} & \colhead{redshift} & \colhead{f(H$\alpha$)} & \colhead{f([N~II]$\lambda$6585)} & \colhead{f([S~II]$\lambda$6718)} & \colhead{f([S~II]$\lambda$6732)}}
\startdata
SGAS J085137$+$333114           &   $1.69320\pm 0.00020$ &   $247\pm27$ &  $67\pm10$   &  $36 \pm 8$  &  $25\pm31$   \\
SGAS J091538$+$382658           &   $1.50200\pm 0.00002$ &   $170\pm5$  &  $13\pm10$   &  \nodata     &  \nodata     \\
SGAS J103842$+$484926           &   $2.19678\pm 0.00001$ &  $1220\pm8$  &  $40\pm11$   &  $68 \pm 9$  &  $67\pm17$   \\
%
SGAS J103844$+$484914           &                        &              &              &              &              \\
...knot1                        &   $0.96540\pm0.00010$  &    $63\pm11$ &  $11\pm3$    &  \nodata     & \nodata      \\
...knot2                        &   $0.96457\pm0.00006$  &   $120\pm9$  &  $17\pm5$    &  \nodata     & \nodata      \\
...total                        &   $0.96479\pm0.00005$  &   $183\pm14$ &  $28\pm6$    &  \nodata     & \nodata      \\ 
SGAS J111124$+$140901           &   $2.13855\pm0.00002$  &   $519\pm7$  &  $34\pm7$    &  $15 \pm 7$  &   $26\pm6$   \\
%
SGAS J113809$+$275445           &                        &              &              &              &              \\
...knot1                        &   $1.33224\pm0.00009$  &    $21\pm2$  &  $1.3\pm1.3$ &  $3.9\pm1.8$ &  $8\pm4$     \\
...knot2                        &   $1.33240\pm0.00010$  &    $13\pm2$  & $-0.9\pm1.2$ &  \nodata     & \nodata      \\
...total 	                &   $1.33231\pm0.00007$  &    $34\pm3$  &  $0.4\pm1.8$ &  \nodata     & \nodata      \\               	        
%
SGAS J113809$+$275439           &                        &              &               &              &              \\
...knot1                        &   $0.90905\pm0.00001$  &   $127\pm2$  &  $4.4\pm 1.9$ &  $14\pm2.6$  &  $5\pm1$     \\
...knot2                        &   $0.90907\pm0.00001$  &    $87\pm2$  &  $2.6\pm 1.7$ &  $8 \pm 2$   &  $5\pm1$     \\
...total                        &   $0.90906\pm0.00001$  &   $214\pm3$  &  $7.0\pm 2.5$ &  $22 \pm 3$  &  $10\pm1.4$  \\
SGAS J120924$+$264052           &   $1.02057\pm0.00002$  &    $51\pm2$  &  $6.7\pm1.3$  &  \nodata     & \nodata      \\
SGAS J134334$+$415509           &   $2.09224\pm0.00005$  &   $155\pm10$ &   $23\pm4$    &  $19 \pm 4$  & $4\pm6$      \\
SGAS J142039$+$395454           &   $2.16256\pm0.00008$  &   $216\pm13$ &   $34\pm11$   &  \nodata     & \nodata      \\
%
%
\enddata
\tablecomments{Flux units are $10^{-17}$~erg~s$^{-1}$~cm$^{-2}$. For objects where we separately extract discrete ``knots'' or images, we report measurements for each component, as well as the summed fluxes over all components, and the weighted mean redshift and the uncertainty in the mean. Only [S~II] detections are included.
}
\end{deluxetable*}

\section{Physical Properties}
\label{sec:param}
\subsection{Lensing mass models}
\label{subsec:massmodel}
A crucial component of any study of a lensed galaxy is the construction of a mass model for the foreground galaxy cluster to derive the lensing magnification and demagnify the observed physical properties back to the source plane of the galaxy. A detailed description of the modeling process for the 8 lens systems considered in this paper will be given elsewhere (K. Sharon et al. 2012, in preparation). Lensing models for SGAS J091538$+$382658 and SGAS J134334$+$415509 can also be found in \cite{Bayliss2010}. To summarize, the lensing models are constructed using the publicly-available software {\tt LENSTOOL} \citep{Jullo2007} with MCMC minimization in the source plane. The clusters are typically modeled with a Navarro, Frenk \& White (1996) profile or a Pseudo-Isothermal Ellipsoid Mass Distribution (PIEMD) to represent the cluster mass. Additional PIEMDs, scaled with luminosity, represent the lensing effect from cluster-member galaxies (see Limousin et al. 2007). The redshifts and positions of the multiply-imaged lensed background galaxies are used as constraints, and when available, the velocity dispersion of the cluster and low-uncertainty photometric redshifts of secondary arcs are used as priors. 

The source plane reconstruction is done by ray-tracing the pixels in the observed frame through the lensing potential of the best-fit lens model. The ratio of the observed image size and the size of the model-reconstructed source provides an estimate of the magnification. The magnification uncertainty is derived from a simulation in which many lens models are computed, with parameters drawn from a set of model parameters from steps in the MCMC that correspond to 1~$\sigma$ uncertainty in the parameter space. The magnifications and their associated uncertainties for the sample of lensed galaxies are noted in Table~\ref{tab:param}.

\subsection{Stellar mass}
\label{subsec:mass}
Stellar masses are determined by fitting spectral energy distribution (SED) models to the available broadband photometry. Following the recipe presented in \cite{me2010}, we use the newest Bruzual and Charlot SED models (CB07, see \cite{Bruzual2003}) with a Chabrier initial mass function \citep{Chabrier2003} and a Calzetti dust extinction law \citep{Calzetti2000}. The metallicity is fixed to the value derived from the rest-frame optical spectroscopy. To avoid unphysically young, luminosity-weighted solutions, the age of the stellar population is required to be $>50$~Myr, which roughly corresponds to the dynamical timescale at $z\sim2$ \citep{Erb2006c}. 
We have carried out the SED fitting with constant star formation (CSF) models as well as a range of exponentially declining star formation histories (SFH), parametrized as $SFR(t)\sim e^{-t/\tau}$ with $\tau$=10, 50, 100, 200, 500, 1000, 2000 and 5000\,Myr. The data generally can not constrain the shape of the SFH; we found that most SFHs provide acceptable fits for most sources. However, the choice of SFH can have a large influence on the stellar population parameters reported by the SED fit, especially the dust extinction and current star formation rate. Recently, \cite{Wuyts2011} have shown that when using exponentially declining SFHs, a minimum e-folding time $\tau=3$~Gyr is required to match SED derived SFRs to SFRs derived from the bolometric UV+IR luminosity. They find that although shorter e-folding times often report lower $\chi^2$ values, these models will underpredict the current star formation rate and underestimate the age of the stellar population. We similarly find that the shortest e-folding times $\tau=10-50$\,Myr are often preferred in a least-squares sense for our lensed galaxies, but the inferred current SFRs are much lower than the SFRs we derive from the H$\alpha$ emission flux (see \S\ref{subsec:sfr}). Since for the relatively young ages of our lensed galaxies (50-300\,Myr), there is little difference between a CSF model and an exponentially declining SFH with $\tau=3$~Gyr, we choose to limit our SED fits to CSF models. We conservatively include the influence of a declining SFH in the systematic uncertainties of the stellar population parameters. Overall, stellar mass is the most robust parameter that can be derived from SED fitting, with an average fractional uncertainty of $\sigma_{M_*}/M_*=0.4$ \citep{Shapley2005, Wuyts2007}. 

One important limitation of single component constant or declining SFHs is their insensitivity to underlying old stellar populations. For a subsample of bright lensed galaxies with similar properties to the sample studied in this paper, we have shown in \cite{me2012} that the contribution of an underlying old stellar population to the galaxy's stellar mass is limited to less than twice the mass derived from a single component SFH. We expect a similar result here and conclude that the stellar mass derived from single-component SED fitting provides a reasonable estimate of the total stellar mass of the lensed galaxies.
\npar
The best-fit values of the stellar mass are reported in Table~\ref{tab:param}. They have been corrected for the lensing magnification. The statistical uncertainties on the stellar mass are derived from 1000 mock realizations of the observed SEDs consistent with the photometric uncertainties. We include a  systematic uncertainty related to the choice of SFH by repeating the fits to the 1000 mock realizations of the observed SED with a range of exponentially declining SFHs.

\begin{deluxetable*}{llllllll}
\scriptsize
\tablewidth{0pc}
\tablecaption{Physical properties of the sample \label{tab:param}}
\tablehead{
\colhead{source} &  \colhead{$\mu$}  & \colhead{$\log(M_*/\mathrm{M}_\odot)$} & \colhead{$SFR_{SED}$}         & \colhead{$E(B-V)$} & \colhead{aperture}   & \colhead{$SFR_{H\alpha}$}     & \colhead{$12+\log(O/H)$} \\
\colhead{}       &  \colhead{}       & \colhead{}                             & \colhead{M$_\odot$~yr$^{-1}$}   & \colhead{}              & \colhead{correction} & \colhead{M$_\odot$~yr$^{-1}$} & \colhead{}} 
\startdata
SGAS J085137$+$333114 & 27$^{+8}_{-5}$   & 9.5$^{+0.2}_{-0.2}$ & 32$^{+12}_{-9}$    & 0.33$\pm$0.14 & 2.7$\pm$0.4 & 60$^{+33}_{-30}$     & 8.87$^{+0.08}_{-0.07}$  \\
SGAS J091538$+$382658 & 25$^{+36}_{-8}$  & 8.4$^{+0.8}_{-0.2}$ & 4.0$^{+6}_{-2}$    & 0.17$\pm$0.12 & 2.0$\pm$0.6 & 15$^{+23}_{-9}$      & 8.42$^{+0.28}_{-0.30}$  \\
SGAS J103842$+$484926 & 9$^{+5}_{-1}$    & 9.9$^{+0.3}_{-0.1}$ & 183$^{+106}_{-36}$ & 0.27$\pm$0.02 & 2.9$\pm$0.7 & 1530$^{+930}_{-410}$ & 8.11$^{+0.11}_{-0.11}$  \\
SGAS J103844$+$484914 & 9$^{+2}_{-2}$    & 9.0$^{+0.3}_{-0.3}$ & 11$^{+4}_{-4}$     & 0.14$\pm$0.11 & 3.2$\pm$0.9 & 22$^{+11}_{-11}$     & 8.66$^{+0.08}_{-0.08}$  \\
SGAS J111124$+$140901 & 14$^{+17}_{-10}$ & 9.2$^{+0.6}_{-0.4}$ & 6$^{+9}_{-6}$      & 0.05$\pm$0.06 & 2.0$\pm$0.6 & 130$^{+170}_{-110}$  & 8.37$^{+0.07}_{-0.07}$  \\
SGAS J113809$+$275445 & 12$^{+2}_{-2}$   & 9.2$^{+0.3}_{-0.3}$ & 37$^{+8}_{-8}$     & 0.30$\pm$0.16 & 4.1$\pm$0.9 & 14$^{+8}_{-8}$       & 8.11$^{+0.14}_{-0.15}$  \\
SGAS J113809$+$275439 & 14$^{+2}_{-3}$   & 8.7$^{+0.1}_{-0.1}$ & 7$^{+2}_{-2}$      & 0.17$\pm$0.11 & 2.9$\pm$0.8 & 14$^{+7}_{-7}$       & 7.67$^{+0.70}_{-0.56}$  \\
SGAS J120924$+$264052 & 58$^{+6}_{-6}$   & 8.6$^{+0.3}_{-0.3}$ & 4$^{+1}_{-1}$      & 0.22$\pm$0.14 & 3.5$\pm$1.2 & 2$^{+1}_{-1}$        & 8.60$^{+0.07}_{-0.07}$  \\
SGAS J134334$+$415509 & 40$^{+4}_{-5}$   & 9.4$^{+0.2}_{-0.2}$ & 13$^{+7}_{-7}$     & 0.11$\pm$0.05 & 3.5$\pm$1.4 & 29$^{+13}_{-13}$     & 8.65$^{+0.07}_{-0.06}$  \\
SGAS J142039$+$395454 & 7$^{+2}_{-1}$    & 9.9$^{+0.2}_{-0.2}$ & 135$^{+63}_{-53}$  & 0.28$\pm$0.05 & 3.7$\pm$0.4 & 440$^{+150}_{-110}$  & 8.67$^{+0.12}_{-0.11}$  \\
\enddata
\tablecomments{Columns are source name, approximate lensing magnification, intrinsic stellar mass derived from the SED fit, intrinsic SFR derived from the SED fit, reddening derived from the SED fit, slit aperture correction, intrinsic SFR derived from the dust-corrected H$\alpha$ emission, and metallicity derived from the [N~II]/H$\alpha$ ratio as calibrated by \cite{Maiolino2008}}
\end{deluxetable*}


\subsection{Star formation rate}
\label{subsec:sfr}
\begin{figure}
\center
\includegraphics[width=0.45\textwidth]{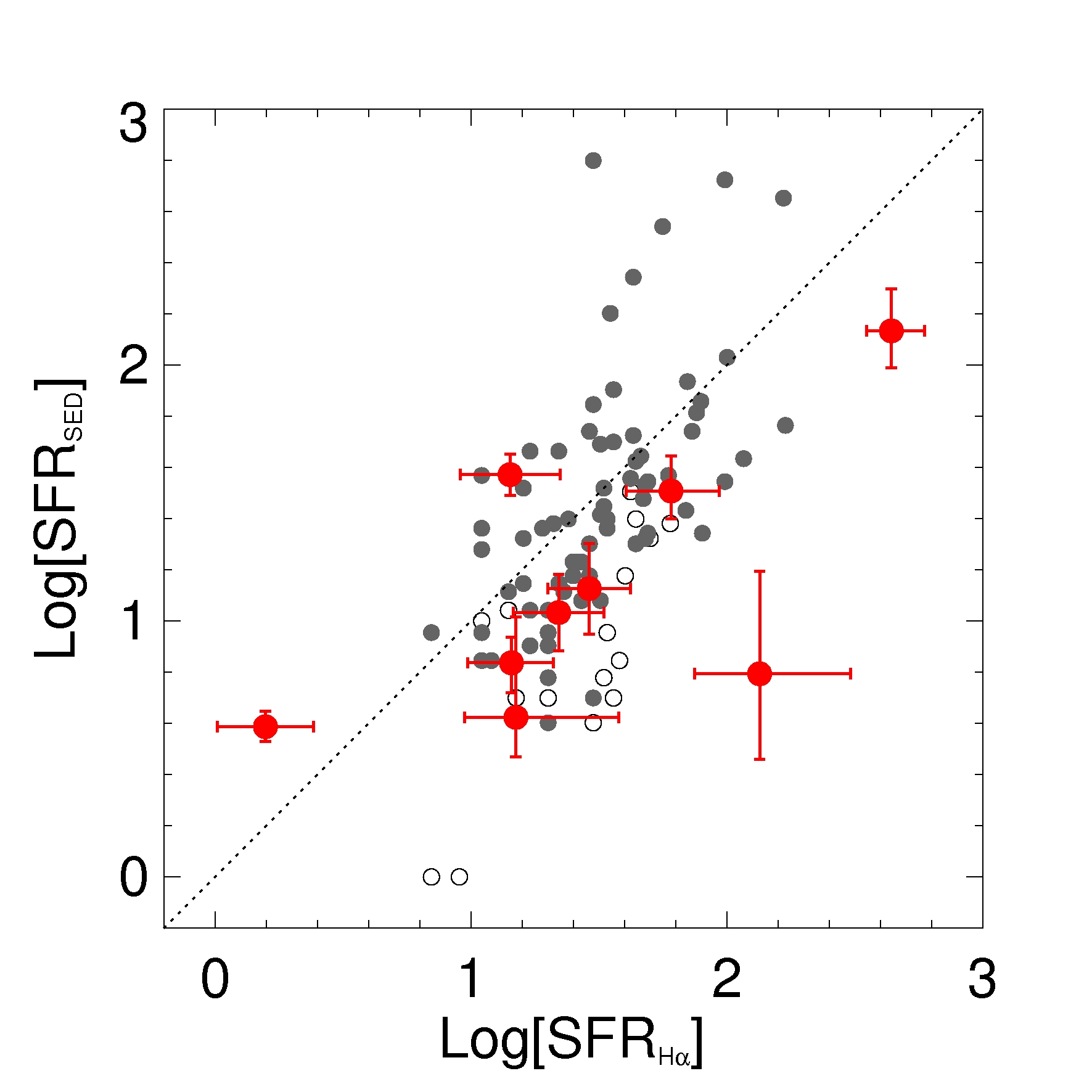}
\caption{Comparison between the SFR derived from the SED fit and the SFR derived from the H$\alpha$ emission. The gray datapoints correspond to the sample of UV-selected galaxies at $z\sim2$ studied by Erb et al.~(2006b,2006c). SED models that asssume an exponentially declining SFH are distinguished as open circles. \label{fig:sfr}}
\end{figure}

Table~\ref{tab:param} reports the current star formation rates for our lensed galaxies, corrected for the lensing magnification, as derived from the range of best-fit SED models. We compare these to star formation rates estimated from the extinction-corrected H$\alpha$ emission flux. For the latter, we have to aperture correct the flux received through the narrow NIRSPEC slit to the total intrinsic flux. To this end, we smooth the $g$-band image of the source with a Gaussian to mimick the NIRSPEC seeing conditions. The $g$-band image is used because the rest-frame UV emission can be expected to most closely match the H$\alpha$ emission. From the smoothed $g$-band image, we measure the fraction of the total flux (within the 2*FWHM photometric aperture) that falls inside the NIRSPEC slit. The correction factors are reported in Table~\ref{tab:param}. The uncertainties on this aperture correction are estimated from the uncertainty in the seeing (which was quite variable during the observations, see \S\ref{subsec:spec}), and a pointing error up to 0.25\arcsec. 
\npar
We derive a star formation rate from the aperture corrected H$\alpha$ flux using the relation from \cite{Kennicutt1998}, converted to the Chabrier IMF by dividing by a factor 1.7. We correct for extinction using the Calzetti dust extinction law and the reddening $E(B-V)$ derived from the best-fit SED models. We assume that the reddening of the nebular emission lines is the same as the reddening of the stellar light as derived from the SED fit. 
\npar
The H$\alpha$ star formation rates are corrected for the lensing magnification and reported in Table~\ref{tab:param}. In Figure~\ref{fig:sfr} we compare $SFR_{H\alpha}$ to $SFR_{SED}$. The plot also shows the sample of UV-selected star-forming galaxies at $z\sim2$ studied in Erb et al.~(2006b,2006c). They use a similar strategy of assuming constant star formation histories for their SED fits (except when exponentially declining models produce a much improved fit - those galaxies are shown with open symbols in Figure~\ref{fig:sfr}), and dust correcting the H$\alpha$ flux with $E(B-V)$ derived from the SED. 
Our lensed galaxies are broadly consistent with their sample, and show a similar scatter of 0.3~dex in the relation between $SFR_{H\alpha}$ and $SFR_{SED}$. The open symbols in this plot illustrate how SED models with exponentially declining SFHs with short e-folding times can significantly underestimate the current SFR, as discussed in \S\ref{subsec:mass}.
\npar
It is worth mentioning that the lensing magnification of our galaxies introduces an additional source of uncertainty in $SFR_{H\alpha}$. When we derive the total intrinsic H$\alpha$ flux by aperture correcting the flux to the whole galaxy and then correcting for the average magnification, we implicitly assume that the magnification is uniform across the galaxy. If the area covered by the slit is more highly magnified than the rest of the galaxy, this procedure will overestimate the intrinsic H$\alpha$ flux and vice versa. We can avoid this uncertainty by mapping both the galaxy and the slit back to the source plane, and measuring the aperture correction factor intrinsically in this plane (K. Sharon et al.~2012, in preparation). Given the overall agreement with the non-lensed sample from Erb et al.~(2006b,2006c) seen in Figure~\ref{fig:sfr}, we do not expect spatial magnification variations to cause a significant bias in $SFR_{H\alpha}$ for our lensed galaxies.


\subsection{Electron density}
\label{subsec:ne}
The doublet ratio of [S~II]$\lambda$6718-$\lambda$6732 constrains the electron density of the nebular gas. We use the IRAF task stsdas.analysis.nebular.temden to calculate $n_e$ for the four lensed galaxies where the lines are sufficiently well detected, assuming an electron temperature of $T_e=10^4$~K. The results are reported in Table~\ref{tab:ne}. We combine these new measurements of electron density with results from the literature for three lensed galaxies from our previous work \citep{Rigby2011, me2012}, two lensed galaxies from \cite{Hainline2009}, and one lensed galaxy from \cite{Bian2010}. The measurements span the full range of plausibility from the low density limit to $n_e>10^4$~cm$^{-3}$, which implies a huge range in physical conditions in the nebular regions of star-forming galaxies at $z\sim2$. There is no evidence for the existence of an ``average'' electron density. This issue should be investigated further by obtaining density measurements for a larger sample of galaxies, as well as deeper spectra to more precisely constrain densities for galaxies with existing data, since most of the current measurements are quite uncertain.

\begin{deluxetable}{llll}
\scriptsize
\tablewidth{0pc}
\tablecaption{Electron density measurements \label{tab:ne}}
\tablehead{
\colhead{source} &  \colhead{z} & \colhead{$n_e$ (cm$^{-3}$)} & \colhead{reference}} 
\startdata
SGAS J103842$+$484926 & 2.20 & $600^{+1600}_{500}$        & this work \\
SGAS J111124$+$140901 & 2.14 & $4500^{+\infty}_{-3550}$   & this work \\
SGAS J113809$+$275445 & 1.33 & $16000^{+\infty}_{-14750}$ & this work \\
SGAS J113809$+$275439 & 0.91 & low $n_e$ limit            & this work \\
RCSGA0327             & 1.70 & $235^{+30}_{-25}$          & Rigby2011 \\
SGAS1527              & 2.76 & $400^{+260}_{-230}$        & Wuyts2012 \\
SGAS1226              & 2.92 & low $n_e$ limit            & Wuyts2012 \\
Clone		      & 2.00 & 1270--2540		  & Hainline2009 \\
Cosmic Horseshoe      & 2.38 & 320-1600                   & Hainline2009 \\
J0900+2234            & 2.03 & $1100^{+5000}_{-700}$      & Bian2010 \\
\enddata
\end{deluxetable}

\subsection{Metallicity}
\label{subsec:z}
The only direct way to determine the abundance of metals in a galaxy's gas reservoir is through a measurement of the electron temperature $T_e$. Higher metallicity results in more cooling via metal emission lines and thus lower temperature. However, the auroral emission lines (transitions from the second lowest to the lowest excited level) required to measure $T_e$, particularly the most widely used line [O~III]$\lambda$4363, become extremely weak at metallicities above $\sim$0.5 solar. Beyond $z\sim1$, the [O~III]$\lambda$4363 line has been detected in only one low-metallicity lensed galaxy \citep{Yuan2009} and an upper limit has been derived for a highly-magnified lensed galaxy of average metallicity \citep{Rigby2011}. This forces us to rely on more easily detected optical emission lines, such as hydrogen recombination lines (H$\alpha$, H$\beta$) and collisionally excited forbidden lines ([N~II]$\lambda$6585, [O~II]$\lambda$3727, [O~III]$\lambda$4959,5007). Photoionization models are used to calibrate flux ratios of these lines as ``strong-line'' metallicity indicators. Unfortunately, these models are subject to significant uncertainties and systematic effects, which result in offsets up to 0.7\,dex between different strong-line indicators. From a large sample of SDSS galaxies, \cite{Kewley2008} present recipes to eliminate these offsets by converting different metallicity indicators to a common calibration scale. 

Using local metallicity calibrations at high redshift assumes that physical conditions such as ionization parameter and electron density that go into the photoionization models on which these calibrations are based, do not evolve with redshift. There is growing evidence suggesting that this is not a valid assumption. High redshift star-forming galaxies are offset with respect to the local star-forming sequence in the diagnostic BPT diagram of [O~III]$\lambda$5007/H$\beta$ versus [N~II]$\lambda$6585/H$\alpha$ \citep{Erb2006a,Liu2008}. Combining measurements for six lensed galaxies at $z\sim2$, we have shown previously that their ionization parameter is roughly twice as high as what is measured for local galaxies \citep{me2012}. A robust in-situ calibration of commonly used strong-line metallicity indicators at $z\sim2$ through direct detection of the [O~III]$\lambda$4363 is crucial to interpret future abundance measurements at high redshift with the next generation of telescopes. The increased flux levels of lensed galaxies could play an important role here. 
\npar
For this analysis, we choose to derive metallicities from the flux ratio of [N~II]$\lambda$6585 to H$\alpha$, the N2 indicator. The close proximity of the lines allows them to be observed in the same filter, thus eliminating flux calibration issues, and makes their ratio insensitive to dust reddening. This is an important point, given that extinction is notoriously difficult to estimate robustly at high redshift. Given these benefits, the N2 indicator has been used most widely for metallicity studies at $z\gtrsim1$. We use the calibration from \cite{Maiolino2008} based on a joint fit to a sample of local low-metallicity galaxies with measurements of the [O~III]$\lambda$4363 auroral line \citep{Nagao2006} and a sample of SDSS DR4 galaxies with metallicities determined from the photoionization models in \cite{Kewley2002}. The calibration has a systematic uncertainty of 0.1~dex. It results in metallicities $\sim0.1$~dex higher than the N2 indicator as calibrated by \cite{Denicolo2002} and $\sim0.2$~dex higher than the calibration by \cite{PP2004}. At high metallicities, the [N~II] line starts to dominate the cooling and saturates, rendering the N2 indicator largely insensitive to metallicity at $12+\log(O/H)\ge9.1$ (in the \cite{Maiolino2008} calibration). 

The metallicity measurements for the lensed galaxies are listed in Table~\ref{tab:param}. The quoted uncertainties only reflect the statistical uncertainties in the [N~II] and H$\alpha$ emission line fluxes. Apart from the systematic uncertainty introduced by the choice of metallicity calibration, one should keep in mind that an integrated metallicity measurement from a long-slit spectrum might not be representative of the true galaxy abundance in the presence of strong metallicity gradients.  

\section{Additional data from the literature}
\label{sec:addata}
To complement our sample of lensed galaxies at $0.9<z<2.5$, we have searched the literature for galaxies within the same redshift range with published measurements of metallicity, stellar mass, and SFR. We limit ourselves to metallicities measured from the N2 indicator. Including other metallicity indicators such as the $R_{23}$ index (based on [O~II]$\lambda$3727, [O~III]$\lambda$4959,5007 and H$\beta$) would increase the sample, but the uncertainties related to the intercalibration of different metallicity indicators would severely complicate the analysis; the conversions presented by \cite{Kewley2008} are derived locally and it is not at all clear given the discussion in \S\ref{subsec:z} that these offsets will be the same at $z\sim2$. The literature sample contains 60 galaxies, including 6 galaxies with upper limits on the [N~II] emission \citep{Shapley2004, Shapley2005, Forster2006, Liu2008, Law2009, Queyrel2009, Wright2009, Queyrel2009, Bian2010, Richard2011, Rigby2011}.

We have converted the values of metallicity, stellar mass and star formation rate reported in these studies to be consistent with the measurement methods described in \S\ref{sec:param}. Stellar mass is the most robust parameter that can be derived from the SED modeling procedure; different assumptions made in terms of the population synthesis models and the range of allowed star formation histories in the various studies will generally not affect the resulting stellar mass by more than a factor of $\sigma_{M_*}/M_*$=0.4 \citep{Erb2006c, Wuyts2007}. We include this as a systematic uncertainty. For the star formation rates, we include a factor of 2 to aperture correct the published H$\alpha$ flux from the spectroscopic slit to the whole galaxy. This is not necessary for the studies that use integral field spectroscopy \citep{Forster2006,Wright2009,Queyrel2009,Law2009}. The SFR is corrected for dust extinction based on the Calzetti dust extinction law and the reddening $E(B-V)$ reported by the SED fit. The reddening depends significantly on the assumptions made in the SED modeling, even though this is not always adequately discussed in the papers. To account for this, we assume a systematic uncertainty of $\sigma_{E(B-V)}/E(B-V)$=0.7. In \cite{Shapley2005} and \cite{Liu2008} the reddening from the SED fit was not reported, but we use their measurement of the Balmer break instead. 

\section{Results}
\label{sec:results}

\subsection{The relation between stellar mass and SFR}
\label{subsec:sfrmass}
To situate our sample of lensed galaxies with respect to the literature sample, it is informative to place them in a diagram of SFR versus stellar mass. Both in the local Universe and at high redshift, star-forming galaxies follow a tight relation between their rate of star formation and their assembled stellar mass, refered to as the main-sequence of star formation \citep{Noeske2007,Elbaz2007,Daddi2007}. This suggests a continuous mode of galaxy growth over extended timescales, possibly fueled by the accretion of gas through cold flows \citep{Dekel2009}. 

\begin{figure}
\center
\includegraphics[width=0.5\textwidth]{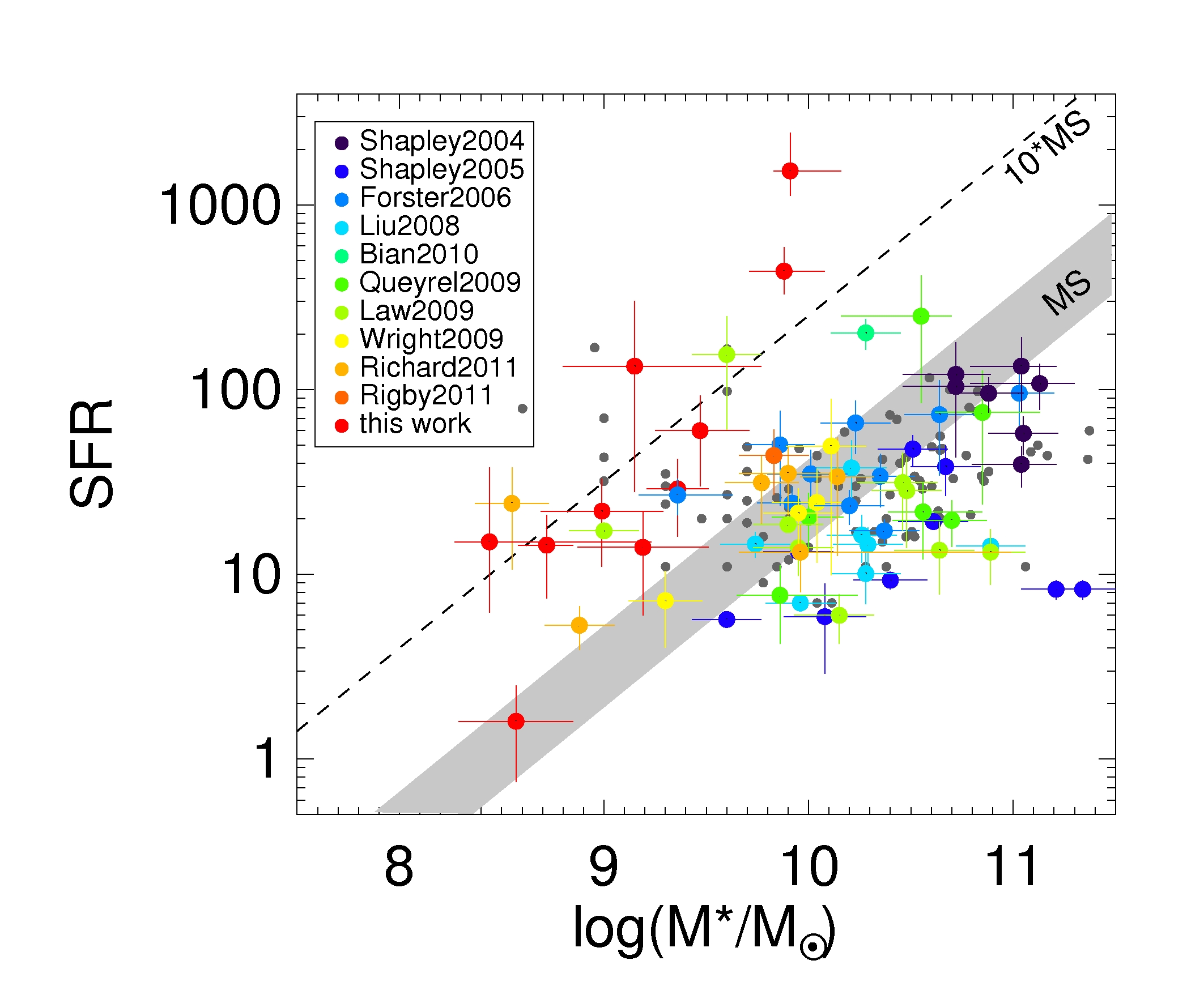}
\caption{The relation between SFR and stellar mass. The literature sample is color-coded by reference; lensed galaxies from this paper are shown in red. The light gray bar indicates the main sequence (MS) of star formation and its standard deviation as identified by \cite{Daddi2007} at $z\sim2$. The dashed line marks sSFRs increased by a factor of 10 compared to the main-sequence. Small gray datapoints correspond to UV-selected galaxies at $z\sim2$ studied by \cite{Erb2006b}.\label{fig:sfrmass}}
\end{figure}

Figure~\ref{fig:sfrmass} shows that the lensed galaxies (red datapoints) probe lower stellar masses than have so far been studied: the median stellar mass of $M_*=10^{9.2}$~M$_\odot$ is a factor of 11 lower than the median stellar mass of the literature sample. The current rate of star formation in the lensed galaxies is high (median $SFR=29$~M$_\odot$~yr$^{-1}$), which can be explained by the fact that we have selected the sample partly based on high rest-frame UV surface brightness (see \S\ref{sec:data}). On average, the lensed galaxies lie a factor of 10 above the main-sequence at $z\sim2$ from \cite{Daddi2007}. This means that their specific star formation rate ($sSFR = SFR/M_*$) is an order of magnitude higher compared to main-sequence star-forming galaxies at $z\sim2$. 

Strongly star-forming outliers to the main-sequence of star formation are typically interpreted as merger-driven starbursts (e.g. Rodighiero et al. 2011). Roughly one third of star-forming galaxies at $z\sim2$ show evidence for a recent merger/interaction in spatially resolved observations of the gas kinematics with near-infrared integral field spectroscopy \citep{Forster2009,Law2009,Wright2009}. These studies have mostly targeted the massive end ($M_* \ge 10^{9.5}$~M$_\odot$) of the main-sequence at $z\sim2$ and there is as of yet no evidence for a correlation between kinematic classification and location in the stellar mass-SFR plane (see Figure 16 in \cite{Forster2009}). Our lensed galaxies offer a unique opportunity to extend these kinematic studies both to lower stellar masses and more strongly star-forming outliers of the main-sequence. 

\begin{figure*}
\center
\includegraphics[width=\textwidth]{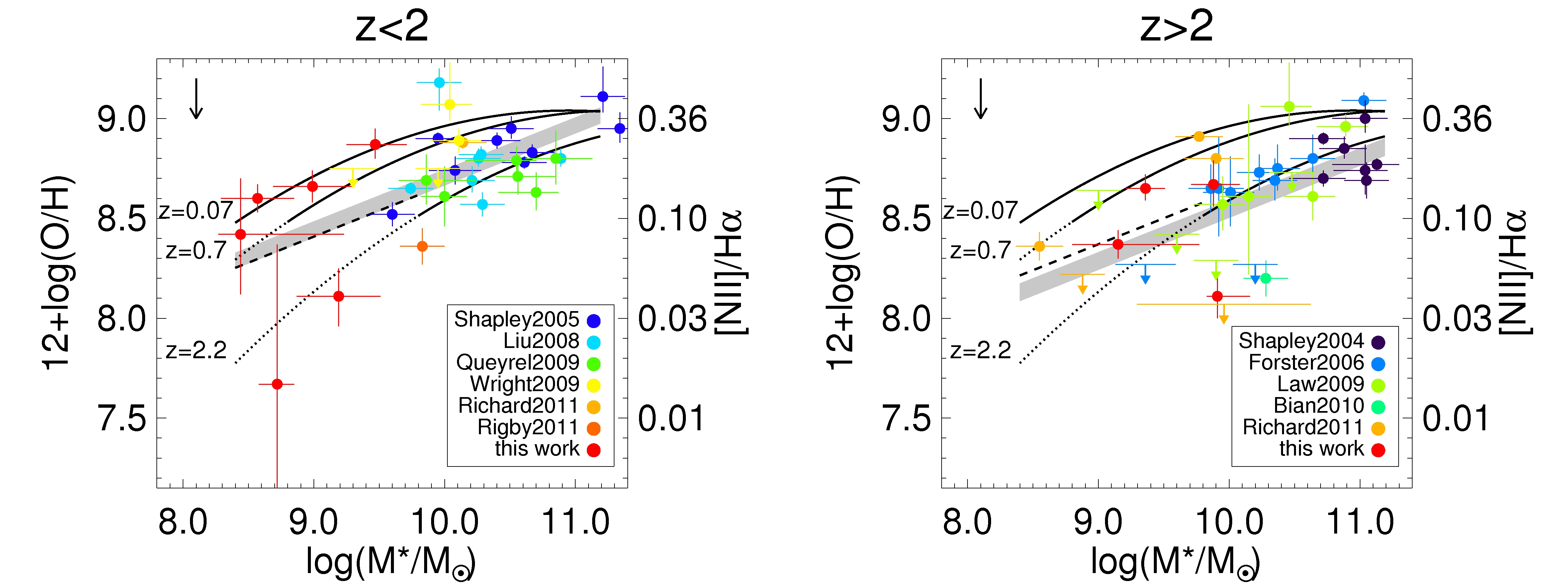}
\caption{Mass-metallicity relation in two redshift bins. The literature sample is color-coded by reference; our lensed galaxies are shown in red. The solid lines show the best-fit polynomial relations at $z=0.07$, $z=0.7$, and $z=2.2$ from \cite{Maiolino2008} based on data from \cite{Tremonti2004}, \cite{Savaglio2005}, and \cite{Erb2006a}; the dotted lines show their extension to stellar masses below the range probed by the respective datasets. The vertical arrow in the top left corner shows a systematic uncertainty of 0.2~dex, based on the fact that the \cite{Maiolino2008} calibration returns metallicities that are $\sim0.2$~dex higher than other N2 indicators (see \S\ref{subsec:z}). The gray areas correspond to 1~$\sigma$ confidence intervals of a linear fit to the data with the slope fixed to the best-fit slope derived for all datapoints combined. The dashed lines show the best linear fits for the subsample at $M_*<10^{9.8}$\,M$_\odot$ in each redshift bin. \label{fig:massmet}}
\end{figure*}
\npar
The lack of galaxies at the low mass - low SFR end of the main-sequence is due to selection effects. Both our sample of lensed galaxies and the literature sample can be described as SFR-limited, since the galaxies are mainly selected based on either UV surface brightness or H$\alpha$ emission. As a comparison, the UV-selected star-forming galaxies at $z\sim2$ studied by \cite{Erb2006b}, shown as small gray datapoints in Figure~\ref{fig:sfrmass}, span a similar range in SFR and stellar mass. 
The full main-sequence only becomes apparent for a near-IR selected sample, which is mass-limited instead of SFR-limited \citep{Daddi2007}.



\subsection{The relation between stellar mass and gas-phase metallicity}
\label{subsec:massmet}
Figure~\ref{fig:massmet} shows the relation between gas-phase metallicity and stellar mass for the combined sample of our lensed galaxies and the data compiled from the literature. We have divided the sample into two redshift bins with roughly equal numbers of sources: 34 galaxies at $z<2$ (median $z=1.4$) and 36 galaxies at $z>2$ (median $z=2.2$). In both redshift bins, the lensed galaxies probe lower stellar masses than have been previously studied. As a reference, we show the best-fit mass-metallicity relations (MZR) derived by \cite{Maiolino2008} (converted to the Chabrier IMF) at $z=0.07$ based on SDSS data from \cite{Tremonti2004}, at $z=0.7$ based on data from \cite{Savaglio2005}, and at $z=2.2$ based on stacked star-forming galaxies from \cite{Erb2006a}. The relations are extended as dotted lines to stellar masses below the range probed by the respective datasets. 

We investigate redshift evolution of the MZR by fitting linear relations to the data in each bin. Offsets in metallicity are more easily quantified with a simple linear fit compared to the polynomial fits used by \cite{Maiolino2008}, especially given the limited number of datapoints and the large scatter. We use the ASURV package (revision 1.3; LaValley et al. 1992) to incorporate both metallicity measurements and upper limits. The dispersion around a linear fit to the data in each redshift bin is 0.23 and 0.25~dex respectively, similar to the dispersion of 0.2~dex found by \cite{Savaglio2005} at $z\sim0.7$, and twice as large as the dispersion of 0.1~dex for SDSS galaxies \citep{Tremonti2004}. 
The gray areas in Figure~\ref{fig:massmet} show the 1~$\sigma$ confidence intervals of the fit when we fix the slope to the best-fit slope derived for all datapoints combined. The y-axis intercepts are 6.10$\pm$0.04 at $z<2$ and 5.94$\pm$0.04 at $z>2$ respectively, which translates into an evolution of $0.16\pm0.06$~dex in the MZR between both redshift bins. The best-fit polynomial relations from \cite{Maiolino2008} find a similar evolution between $z\sim0.7$ and $z\sim2.2$ at high stellar masses.

The lensed galaxies studied in this paper provide the first direct probe of the MZR at $z=1-2$ at low stellar masses. We derive best-fit linear relations for the subsample with stellar mass $M_*<10^{9.8}$~M$_\odot$, which corresponds to the lowest mass bin for which \cite{Erb2006a} was able to measure a metallicity for their stacked sample at $z\sim2.2$. The best linear fit shifts downwards by $0.04\pm0.14$~dex in metallicity from the low to the high redshift bin (dashed lines in Figure~\ref{fig:massmet}). Thus, we find \textit{less} redshift evolution of the MZR at low stellar masses. A larger sample of metallicity measurements for low-mass galaxies is needed to reduce the statistical uncertainties, which are large and formally consistent with no change in redshift evolution from low to high stellar masses. However, our measurements definitely do not agree with the much larger redshift evolution (up to 0.4~dex at $M_*=10^9$\,M$_\odot$) predicted by extrapolations of the polynomial relations from \cite{Maiolino2008} to stellar masses below the measurements at $z=2.2$ from \cite{Erb2006a}. 

\begin{figure}
\center
\includegraphics[width=0.5\textwidth]{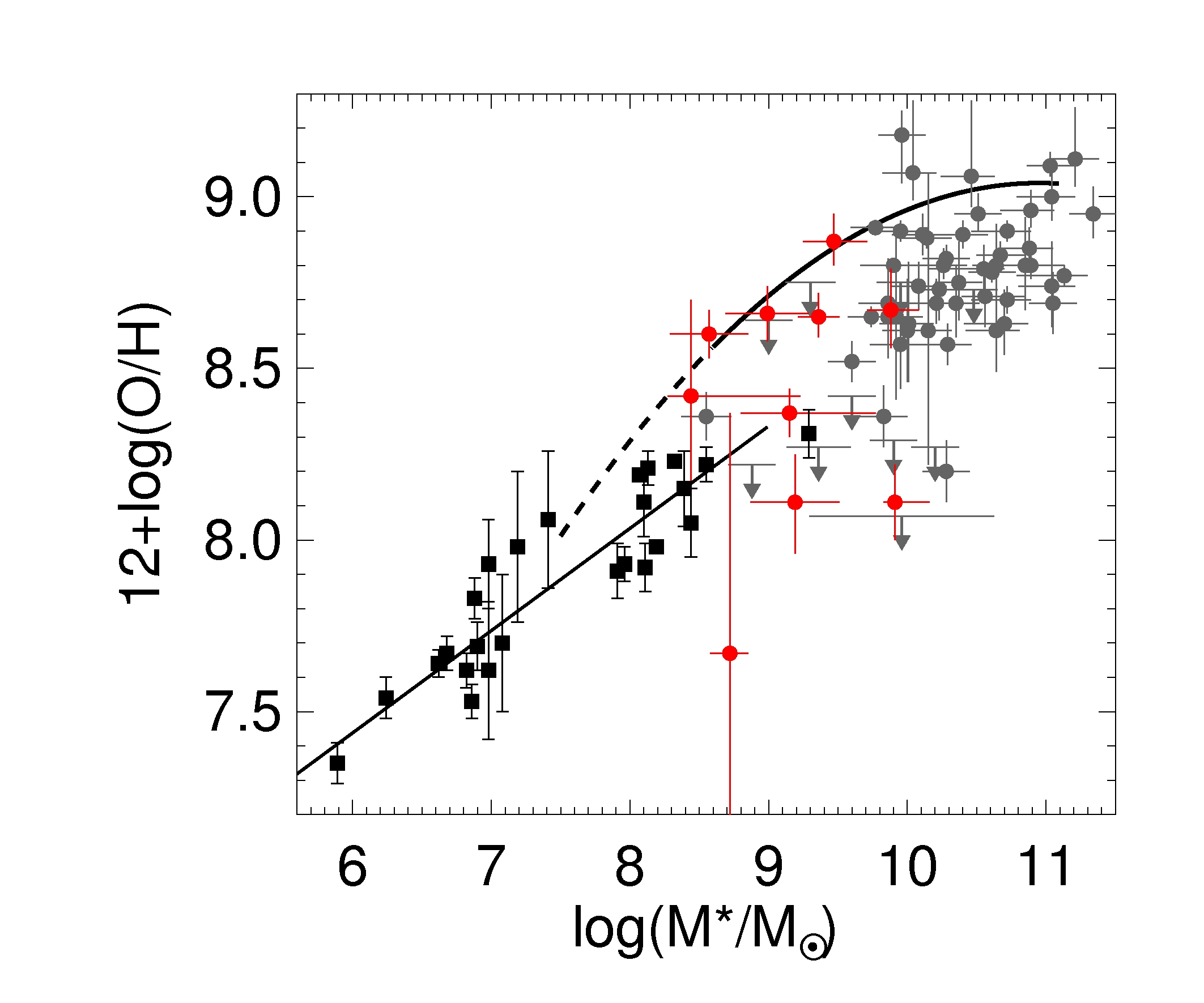}
\caption{Mass-metallicity relation for our sample of lensed galaxies at $0.9<z<2.5$ (red circles), the sample compiled from the literature in the same redshift range (gray circles and upper limits) and the sample of local dwarf galaxies from \cite{Lee2006} (black squares). The linear fit to the dwarf galaxies and the best-fit relation for the SDSS galaxies from \cite{Maiolino2008} are shown as solid black lines. The dashed line corresponds to a linear extension of the SDSS relation to $M_*=10^{7.5}$\,M$_\odot$. \label{fig:massmetdwarfs}}

\end{figure}

\npar
The mass-metallicity relation can be extended to even lower stellar masses when looking at dwarf galaxies. In Figure~\ref{fig:massmetdwarfs} we show results for 25 nearby dwarf galaxies from \cite{Lee2006}. These authors derived stellar masses from the 4.5\,\micron\ luminosity and metallicities from the [O~III]$\lambda$4363 emission line and find a best-fit mass-metallicity relation with a scatter of 0.12~dex, similar to the scatter in the relation for SDSS galaxies \citep{Tremonti2004}. In the region where the sample of dwarf galaxies overlaps with the SDSS sample, \cite{Lee2006} report a metallicity offset of 0.2-0.3~dex and attribute this to an overestimate of the SDSS metallicities at low stellar masses due to the limited central region of nearby galaxies probed by the SDSS fibres and/or an offset between bright-line metallicity indicators and the [O~III]$\lambda$4363 emission. 
Figure~\ref{fig:massmetdwarfs} shows an overall agreement between our lensed galaxies at $0.9<z<2.5$ and the nearby dwarf galaxies with comparable stellar masses. However, given the different methods used to derive both stellar mass and metallicity, there could be significant systematic uncertainties. A direct comparison requires measurements of the [O~III]$\lambda$4363 line at high redshift, which to date has been done for only one low-metallicity galaxy (Yuan \& Kewley 2009, see \S\ref{subsec:z}).

\subsection{The fundamental metallicity relation}
\label{subsec:fmr}
\begin{figure*}
\center
\includegraphics[width=\textwidth]{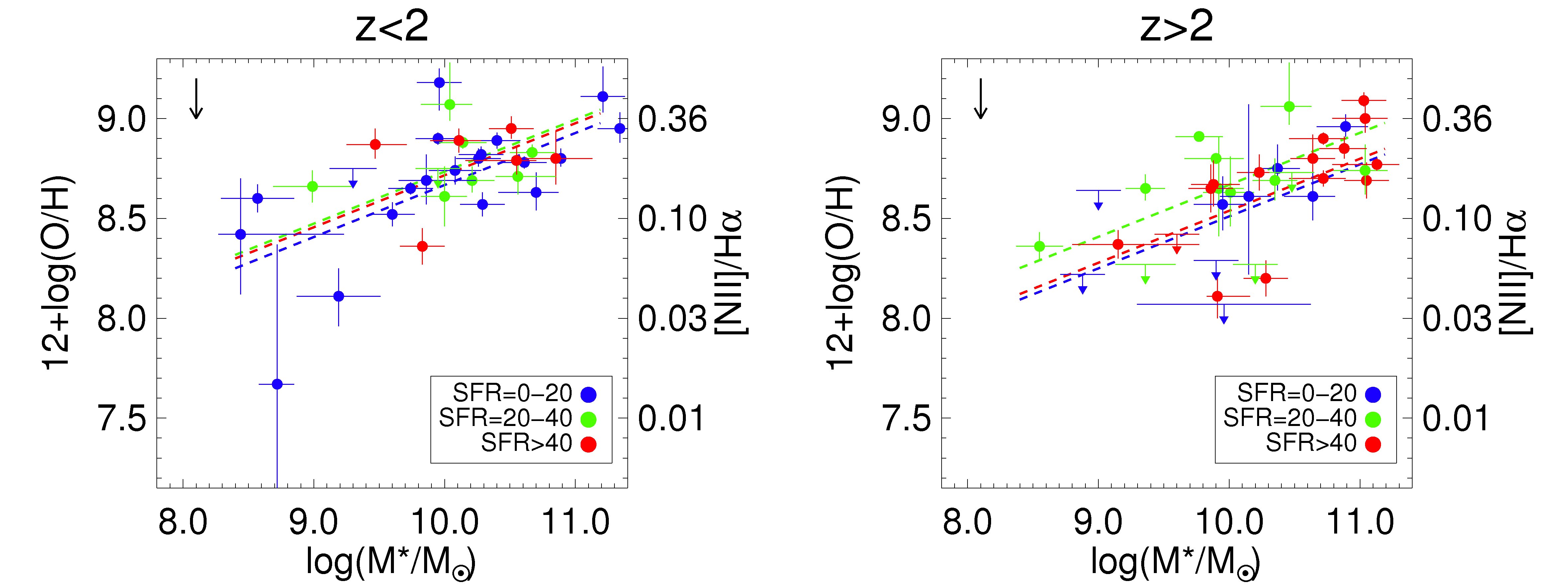}
\caption{Mass-metallicity relation in two redshift bins, color-coded by SFR. The dashed lines show the best-fit linear relations for each SFR bin, keeping the slope fixed to the best-fit slope for all datapoints combined (see \S\ref{subsec:massmet}). The vertical arrow in the upper left corner corresponds to the systematic uncertainty in the metallicity calibration, as in Figure~\ref{fig:massmet}. \label{fig:massmetsfr}}
\end{figure*}

Given that star formation is responsible for the build up of the stellar mass and metal content of galaxies, it is instructive to study the dependence of the mass-metallicity relation on star formation rate. Several studies have reported a correlation between metallicity and (s)SFR for SDSS galaxies, where at a fixed stellar mass, more actively star-forming galaxies show lower gas-phase metallicities \citep{Ellison2008, Mannucci2010, Laralopez2010, Yates2011}.  
We examine the dependence of metallicity on SFR for our lensed galaxies and the sample from the literature in Figure~\ref{fig:massmetsfr}. For each redshift bin, we divide the combined sample into three bins of SFR, chosen to include a roughly equal number of sources in each bin, $SFR < 20$~M$_\odot$~yr$^{-1}$, $20 < SFR < 40$~M$_\odot$~yr$^{-1}$, and $SFR > 40$~M$_\odot$~yr$^{-1}$. Figure~\ref{fig:massmetsfr} shows no evidence for a correlation between metallicity and SFR at fixed stellar mass. We quantify this by computing best-fit linear relations for each redshift and SFR bin, keeping the slope fixed as in \S\ref{subsec:massmet}. For this sample, the highest SFRs do not correspond to the lowest metallicities. Given the 1~$\sigma$ uncertainties of $\sim$0.09~dex on the y-axis intercepts of the fits, their separation is not statistically significant; for each redshift bin, the best-fit linear relations for the three SFR bins are consistent. A recent study of star-forming galaxies at $z\sim1.4$ with Subaru, limited to high stellar masses of $M_*>10^{9.5}$\,M$_\odot$, similarly found only very weak evidence for a correlation between metallicity and SFR at fixed stellar mass \citep{Yabe2011}. 

\begin{figure}
\center
\includegraphics[width=0.5\textwidth]{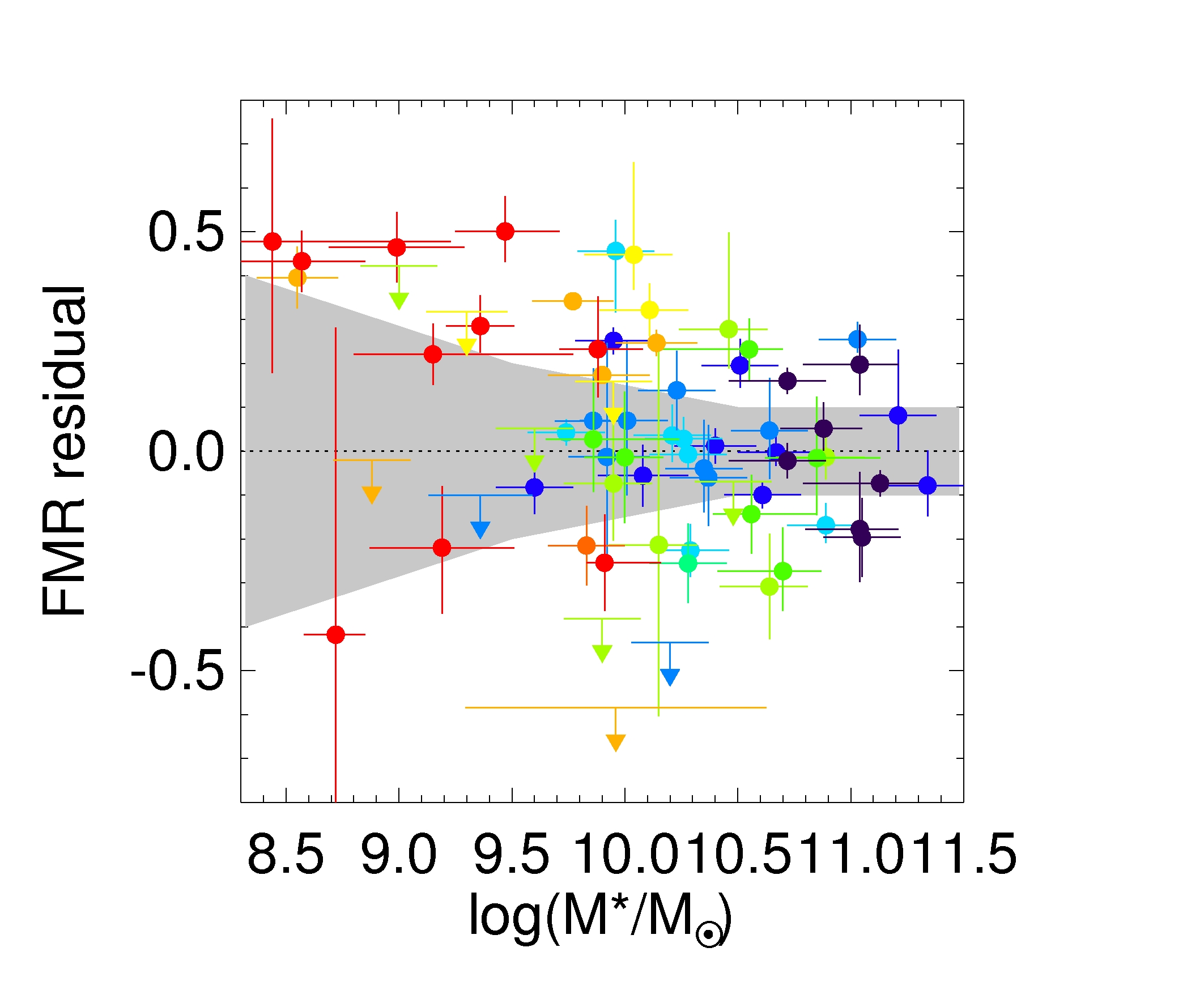}
\caption{Residual from the \cite{Mannucci2010} FMR. The literature sample is color-coded by reference as in Figure~\ref{fig:sfrmass} and Figure~\ref{fig:massmet}; our lensed galaxies are shown in red. The gray area shows the 1~$\sigma$ uncertainty in the FMR for SDSS galaxies (Mannucci et al. 2010, 2011) \label{fig:fmrres}}
\end{figure}

\begin{figure}
\center
\includegraphics[width=0.5\textwidth]{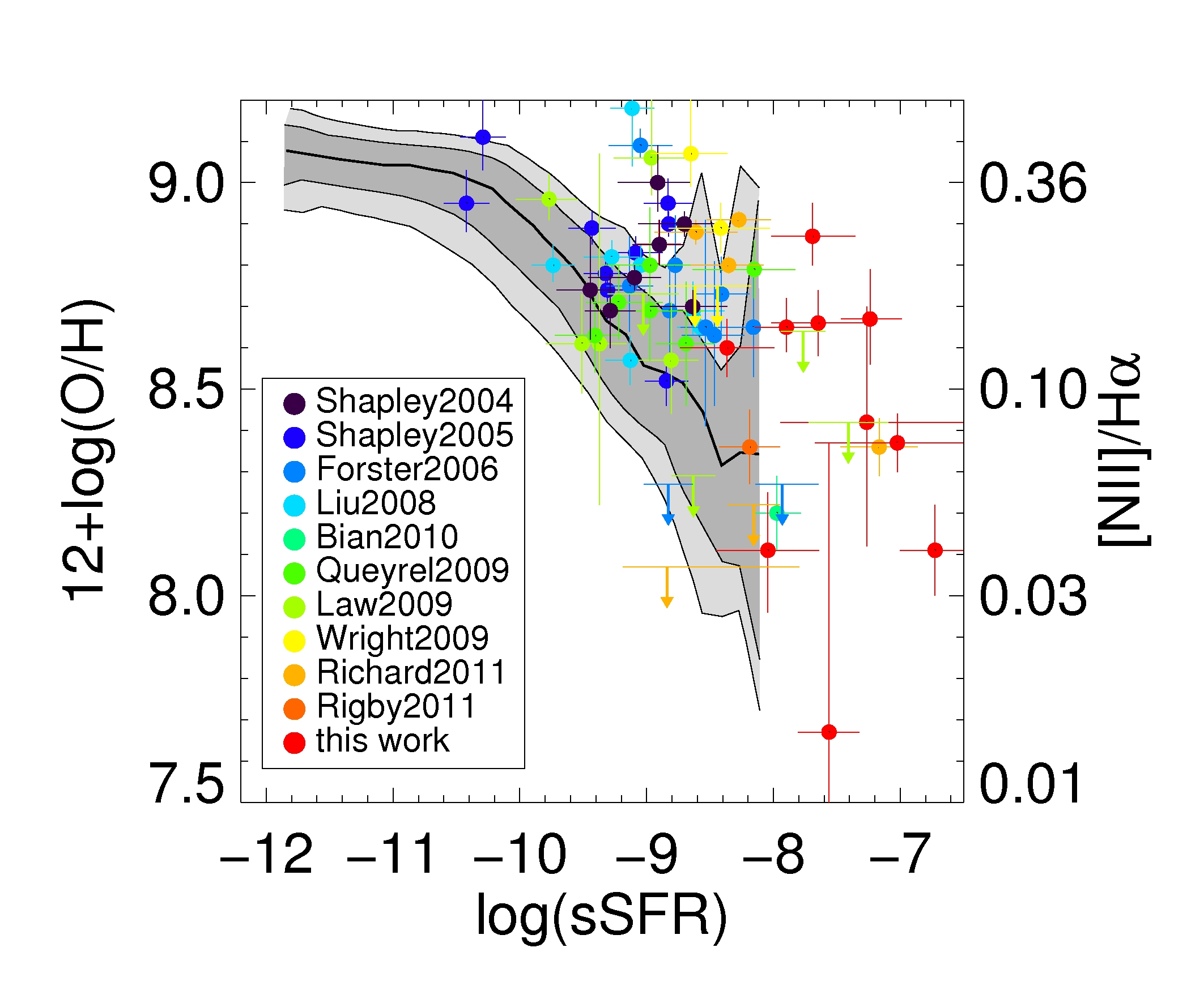}
\caption{Metallicity as a function of specific star formation rate, based on Figure 2 in \cite{Mannucci2011}. The gray areas show 68\% and 90\% confidence intervals for SDSS galaxies. The literature sample is color-coded by reference as in Figure~\ref{fig:sfrmass}, Figure~\ref{fig:massmet}, and Figure~\ref{fig:fmrres}; our lensed galaxies are shown in red. \label{fig:fmr}}
\end{figure}

It is worth pointing out that the SFR dependence of the metallicity reported by \cite{Mannucci2010} applies to SDSS galaxies with low SFRs ranging from 0.04 to 6.0~M$_\odot$~yr$^{-1}$. The high redshift galaxies studied in this paper are much more actively star-forming, with SFRs ranging from 1.7 to 1530~M$_\odot$~yr$^{-1}$. It would be very instructive, though observationally challenging, to look for a correlation between metallicity and SFR in less active high-redshift galaxies with a range of SFRs comparable to SDSS galaxies.
\npar
\cite{Mannucci2010} have proposed a fundamental metallicity relation (FMR) where galaxy abundance depends on both SFR and stellar mass, which reduces the intrinsic scatter in the local MZR to 0.06~dex. Based on median values of the available literature data, they find that star-forming galaxies out to $z=2.5$ follow the same FMR. They explain the observed redshift evolution of the mass-metallicity relation through the selection of galaxies with progressively increasing SFRs at higher redshift. In a follow-up paper, \cite{Mannucci2011} extend the FMR to lower stellar masses, from $M_*>10^{9.2}$\,M$_\odot$ to $M_*>10^{8.3}$\,M$_\odot$, based on $\sim$1400 low-mass SDSS galaxies, $\sim1$\% of the total sample of SDSS-DR7 galaxies used in \cite{Mannucci2010}. The scatter in the relation increases dramatically, up to 0.35~dex at $M_*=10^{8.5}$\,M$_\odot$. Figure~\ref{fig:fmrres} shows the residual from the FMR for our sample of lensed galaxies and the sample from the literature. There is a general agreement at all stellar masses, the median residual is 0.29$\pm$0.28~dex. The gray area in Figure~\ref{fig:fmrres} corresponds to the 1~$\sigma$ uncertainty in the FMR (Mannucci et al. 2010, 2011) -- the high redshift sample generally shows a larger scatter.  
\npar
Figure~\ref{fig:fmr} is based on Figure 2 in \cite{Mannucci2011} and shows metallicity as a function of specfic star formation rate. The gray areas contain 68\% and 90\% of the SDSS galaxies used for their derivation of the FMR. The scatter in the metallicity increases dramatically with sSFR. The high-redshift galaxies studied in this paper are generally consistent with the SDSS confidence interval. The sSFRs of the lensed galaxies are on average an order of magnitude of higher than the literature sample, which generally follows the main-sequence of star formation at $z\sim2$ (see also \S~\ref{subsec:sfrmass}). The lensed galaxies thus probe a new range of high sSFR, not present in the local SDSS sample. In this range, metallicity continues to decrease with sSFR, and the scatter remains substantial.

\section{Discussion}
\label{sec:disc}
As discussed in the introduction to this paper, the correlation between stellar mass and metallicity is a natural consequence of the conversion of gas into stars within galaxies, mitigated by gas exchanges with the environment through inflows and outflows. Since gas flows are hard to observe directly at high redshifts, the observed mass-metallicity relation can indirectly constrain their relative importance for the chemical evolution of galaxies. In this discussion, we limit ourselves to simple analytical models of chemical evolution. This undoubtedly does not capture the full complexity and interplay of the physical processes involved, but we will show that the combination of large scatter and significant systematic uncertainties in the key observables, do not allow constraints on even these simple models.

\subsection{Analytical models}
\label{subsec:an}
\begin{figure*}
\center
\includegraphics[width=\textwidth]{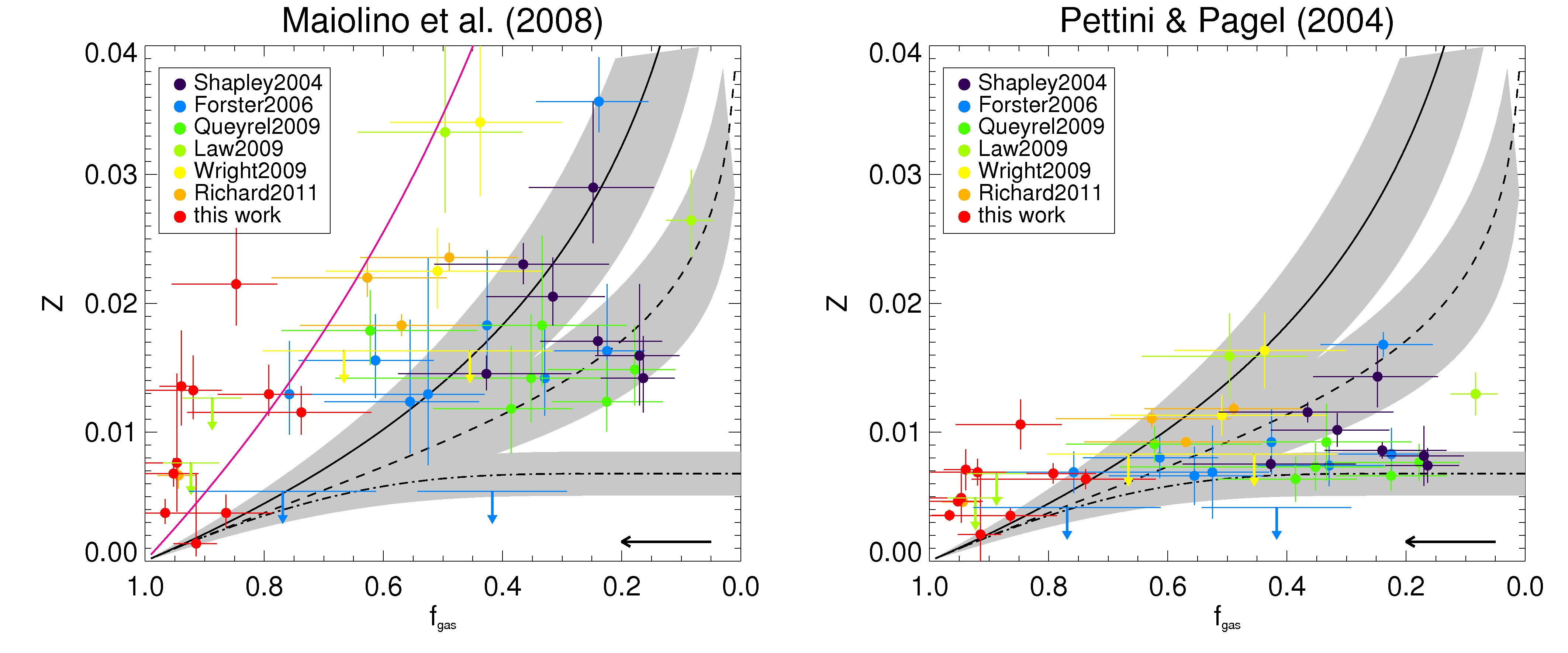}
\caption{Relation between metal mass fraction $Z$ and gas mass fraction $f_{gas}$. We show a closed-box model \textit{(solid line)}, an outflow model with $f_o=2.0$ \textit{(dashed line)}, and a model with outflows at $f_o=2.0$ and inflows at $f_i=f_o+0.95$ \textit{(dot-dashed line)}. In each case, the lines correspond to $y=0.015$ and the gray areas show the confidence intervals that correspond to a range of yields $y=0.010-0.020$. The horizontal arrow shows the systematic uncertainty in $f_{gas}$ from the uncertainty in the slope of the KS law (see \S\ref{subsec:gasmass}). In the left panel, an integrated yield $y=0.05$ is required to obtain a closed-box model that serves as an upper limit to the metal mass fraction as calibrated by \cite{Maiolino2008} (magenta line). The right panel shows the calibration by \cite{PP2004}. \label{fig:zmu}}
\end{figure*}

In the closed-box model of chemical evolution, galaxies do not interact with their environment and the metallicity is fully determined by the galaxy's gas fraction $f_{gas}=M_{gas}/(M_{gas}+M_*)$ and by the integrated stellar yield $y$, the mass of metals produced by massive stars normalized to the stellar mass locked-up in long-lived stars and stellar remnants. 
\begin{equation}
Z=y\ln(f_{gas}^{-1})
\end{equation}
In this equation, $Z$ is the metal mass fraction, the fraction by mass of elements heavier than helium. The integrated yield $y$ is derived from the yields of individual stars, weighted with the adopted initial mass function. Due to remaining uncertainties in stellar evolution and (explosive) nucleosynthesis calculations, the integrated yield is not known to better than a factor of 2. The solid line in Figure~\ref{fig:zmu} shows the closed-box model for $y=0.015\pm0.005$, the appropriate integrated yield for the Chabrier IMF \citep{Finlator2008, Peeples2011}.
Simple analytical extensions of the closed-box model which incorporate inflows and outflows are comprehensively discussed by \cite{Erb2008} -- see also \cite{Pagel1997} and \cite{Spitoni2010}. Outflows are assumed to occur at a constant fraction of the SFR ($f_o$), with the outflowing gas showing the same metallicity as the remaining gas reservoir. In this model, the outflow rate does not depend on the mass of the galaxy. The dashed line in Figure~\ref{fig:zmu} corresponds to a model with $f_o=2.0$, showing a shallower increase of the metallicity with decreasing gas fraction due to the removal of metals by the outflows. To add inflows, the model assumes that gas is accreted at a constant fraction of the SFR ($f_i$) and that the metallicity of the inflowing gas is negligible with respect to the metallicity of the galaxy's gas reservoir. The inflow rates are generally required to be sufficient to roughly replenish the processed gas; 
an accretion rate below 85\% of the gas processing rate fails to reproduce the observed trends of star formation rate with galaxy age at $z\sim2$ \citep{Erb2008}. The dot-dashed curve in Figure \ref{fig:zmu} shows the added effect of inflows at a rate $f_i=f_o+0.95$. The metallicity is seen to asymptote to its equilibrium value $Z_{eq}=y/f_i$ at low gas fractions where a balance between the rate of enrichment from star formation and the rate of dilution from infall of metal-poor gas sets in. Figure \ref{fig:zmu} shows that the uncertainty in the integrated yield creates significant overlap between the different analytical models in the $Z$-$f_{gas}$ plane.


\subsection{Gas fractions}
\label{subsec:gasmass}
A crucial component of the analytical models presented above is the mass of a galaxy's gas reservoir. (Sub)millimeter observations of molecular gas have only recently become feasible for main-sequence star-forming galaxies at $z=1-2$ through significant advances in the sensitivity of the IRAM Plateau de Bure millimeter Interferometer (PdBI; Guilloteau et al. 1992). Additionally, the Atacama Large Millimeter Array (ALMA) will provide unique opportunities for enhanced sensitivity and spatial resolution for molecular gas studies in the near future. Current samples of high-redshift main-sequence star-forming galaxies observed with the PdBI are still small (15 sources at $z=1-1.5$ and 10 sources at $z\sim2.3$) and biased towards the massive end of the main-sequence with $M_* \ge 3 \times 10^{10}$~M$_\odot$ and $SFR \ge 40$~M$_\odot$~yr$^{-1}$ \citep{Tacconi2010, Daddi2010}. 

In the absence of direct measurements of the gas mass, it has become common practice to use the empirical Kennicutt-Schmidt (KS) law to estimate the gas surface density from the SFR density:  $\Sigma_{SFR}\sim\Sigma_{gas}^n$ \citep{Kennicutt1998}. The half-light radius $R$ is typically used to convert between SFR and SFR surface density, such that $M_{gas}=\xi SFR^{1/n} R^{(2n-2)/n}$. The slope $n$ can be calibrated from the current small samples of gas measurements for main-sequence star-forming galaxies at $z=1-2$; results vary from $n=1.17$ \citep{Genzel2010} to $n=1.42$ \citep{Daddi2010}. 

To estimate the gas masses for our sample of lensed galaxies, we compute the half light radius $R$ by mapping the photometric aperture that contains half of the total flux back to the source plane of the galaxies using the lensing model. For the sample from the literature, we use the published half-light radii when available -- no measurements were published for the galaxies taken from \cite{Shapley2005}, \cite{Liu2008}, \cite{Bian2010} and \cite{Rigby2011}, which corresponds to 42\% of the sample. We use the KS law from \cite{Daddi2010} with $n=1.42$ and $\xi=4.98\times10^8$; the best-fit KS law from \cite{Genzel2010} with $n=1.17$ leads to gas fractions that are on average 15\% larger. 

\subsection{Observational constraints on analytical models?}
Figure~\ref{fig:zmu} shows the observed metal mass fractions $Z$, and the gas fractions $f_{gas}$ infered from the KS law for our lensed galaxies and the subsample from the literature for which the half-light radius has been measured. The data span almost the full range of gas fractions and there is no obvious correlation with the metal mass fraction. Theoretically, the phase space to the left of the closed-box model is not allowed. At a given gas fraction, a galaxy can not have a higher metal mass fraction than predicted by the closed-box model, since interactions with the environment through metal-rich outflows and metal-poor inflows will only lower the metallicity. An integrated yield $y=0.05$ is required to obtain a closed-box model that serves as an upper limit to the metallicity for most datapoints (magenta line in the left panel of Figure~\ref{fig:zmu}). Even though the integrated yield is not known to great precision, a value of $y=0.05$ is unphysically high.

The high metal mass fractions in Figure~\ref{fig:zmu} could alternatively suggest that the \cite{Maiolino2008} metallicity calibration is biased high. When applied to SDSS galaxies, the \cite{Maiolino2008} calibration finds a MZR very similar to the result from \cite{Tremonti2004}. This MZR lies towards the high-metallicity end of the 0.7~dex spread in metallicity found by \cite{Kewley2008} for 10 different calibrations of the local MZR (see their Figure~2). In the right panel of Figure~\ref{fig:zmu}, we show metal mass fractions derived from the third order polynomial calibration of the [N~II]/H$\alpha$ ratio from \cite{PP2004}. This returns metallicities that are on average 0.2~dex lower than the \cite{Maiolino2008} calibration (see \S\ref{subsec:z}) and falls in the middle of the range of MZRs from \cite{Kewley2008}. For this calibration, an integrated yield of $y=0.015\pm0.005$ is no longer at odds with most of the data. 

Our lensed galaxies have very high gas fractions of 70-95\% and consistently show larger metal mass fractions than allowed by the closed-box model, even for the \cite{PP2004} metallicity calibration. \cite{Daddi2010} and \cite{Genzel2010} have shown that both local and high-redshift submm galaxies, whose extreme SFRs are powered by merger-driven starbursts, have much higher SFR densities at a fixed gas surface density; their KS law has a similar slope compared to main-sequence star-forming galaxies, but is normalized higher by $\sim0.9$~dex. As seen in \S~\ref{subsec:sfrmass}, the lensed galaxies are outliers of the main-sequence of star formation at $z\sim2$ with 10 times higher specific star formation rates, which is typically attributed to merger-driven star formation. If these galaxies are indeed powered by a recent merger, we are overestimating their gas masses by 20\% on average when using the lower normalization of the KS law for main-sequence star-forming galaxies. This uncertainty can be addressed with either integral field spectroscopy to establish the kinematics of the lensed galaxies or direct measurements of their gas mass.
\npar 
In summary, the current situation does not allow observational constraints on simple analytical models of chemical evolution introduced in \S\ref{subsec:an}. We have identified three sources of systematic uncertainty that need to be addressed. First of all, a robust absolute calibration of the metallicity scale is required. 
We need detections of the [O~III]$\lambda$4363 emission line at high redshift to provide an in-situ direct metallicity calibration via the electron temperature. Until the next generation of telescopes (e.g., JWST) comes online, the increased flux levels of lensed galaxies may provide the only path forward in this respect. Secondly, we need direct measurements of the gas mass for large samples of star-forming galaxies at $z=1-2$, both for main-sequence galaxies and outliers. This is essential to reduce the uncertainty in both the slope and the normalization of the KS law. Even better, a significant galaxy sample with both metallicity and gas mass measurements can avoid the KS law entirely. Lastly, the uncertainty on the integrated yield causes large overlaps between analytical models in the $Z$-$f_{gas}$ plane, limiting our ability to distinguish between them. 



\section{Summary}
This paper presents measurements of the stellar mass, star formation rate and metallicity for 10 lensed galaxies at $0.9<z<2.5$,  identified in the SDSS DR7. We supplement this sample with 60 mostly non-lensed galaxies from the literature in the same redshift range, limited to sources with metallicities derived from the [N~II]/H$\alpha$ flux ratio. Our conclusions are summarized as follows: 
\begin{enumerate}
\item The lensed galaxies probe stellar masses that are on average a factor of 11 lower than have previously been studied in the literature. Their current rate of star formation places them on average a factor of 10 above the main-sequence of star-formation at $z\sim2$, which can be explained by the fact that they were selected for near-IR spectroscopy from a much larger sample of lensed galaxies (M.~D. Gladders et al. 2012, in preparation) partly based on high rest-frame UV surface brightness.
\item We find a redshift evolution of $0.16\pm0.06$~dex in the mass-metallicity relation (MZR) from $z\sim1.4$ to $z\sim2.2$. The polynomial best-fit MZRs from \cite{Maiolino2008} based on data at $z=0.7$ from \cite{Savaglio2005} and stacked galaxies at $z=2.2$ from \cite{Erb2006a} show a similar evolution at high stellar masses. The lensed galaxies provide the first direct constraints on the MZR at $z=1-2$ at stellar masses below $10^{9.8}$~M$_\odot$. Contrary to the extrapolated polynomial best-fit MZRs from \cite{Maiolino2008}, we find \textit{less} redshift evolution at these low stellar masses.  
\item We do not see a correlation between metallicity and SFR for our combined sample, which probes higher SFRs than the SDSS galaxies for which this correlation has previously been reported. We find a general agreement with the local fundamental relation between metallicity, stellar mass and star formation rate (Mannucci et al. 2010, 2011) at all stellar masses, albeit with larger scatter.  
\item We use the Kennicutt-Schmidt law to infer gas masses for the sample and investigate whether we can constrain simple analytical models of chemical evolution with the observed metallicities and gas fractions. An integrated yield $y=0.05$ is required such that the metal mass fractions are bounded by the closed-box model. This is unphysically high, and suggests that the \cite{Maiolino2008} metallicity calibration is biased high. 
\end{enumerate}
\npar
This paper has shown that strongly lensed galaxies offer unique opportunities to extend studies of the mass-metallicity relation to the low-mass end of the galaxy population at $z=1-2$. We are currently obtaining spectroscopy for a larger sample of lensed galaxies to reduce the statistical uncertainties and better constrain the redshift evolution of the relation at low stellar masses. Additionally, we expect the increased flux levels of lensed galaxies to play a crucial role in the path towards reducing the current systematic uncertainties in the metallicity calibration and KS law. 

\begin{acknowledgments}
We thank Andrey Kravtsov and Dawn Erb for helpful discussions. MDG thanks the Research Corporation for support of this work through a Cottrell Scholars award. 

This work includes observations obtained at the Gemini Observatory, which is operated by the Association of Universities for Research in Astronomy, Inc., under a cooperative agreement with the National Science Foundation (NSF) on behalf of the Gemini partnership: the NSF (United States), the Science and Technology Facilities Council (United Kingdom), the National Research Council (Canada), CONICYT (Chile), the Australian Research Council (Australia), Minist\'{e}rio da Ci\^{e}ncia, Tecnologia e Inova\c{c}\~{a}o (Brazil) and Ministerio de Ciencia, Tecnolog\'{i}a e Innovaci\'{o}n Productiva (Argentina).

This work made use of observations made with the Spitzer Space Telescope, which is operated by the Jet Propulsion Laboratory, California Institute of Technology under a contract with NASA. Partial support for this work was provided by NASA through an award issued by JPL/Caltech. 

Data presented in this paper were partly obtained at the W.M. Keck Observatory from telescope time allocated to the National Aeronautics and Space Administration through the scientific partnership with the California Institute of Technology and the University of California. The Observatory was made possible by the generous financial support of the W.M. Keck Foundation. We acknowledge the very significant cultural role and reverence that the summit of Mauna Kea has always had within the indigenous Hawaiian community. We are most fortunate to have the opportunity to conduct observations from this mountain.
\end{acknowledgments}


\end{document}